\documentclass{IEEEtaes}

\usepackage{amsmath}
\usepackage{graphicx}
\usepackage{hyperref}
\usepackage{dsfont}

\usepackage[T1]{fontenc}
\usepackage{graphicx}

\usepackage{subfigure}
\usepackage{epstopdf}

\usepackage{overpic}
\usepackage{amssymb}
\usepackage{amsmath}
\usepackage{amsthm}
\usepackage{array}
\usepackage{color}
\usepackage{url}

\usepackage{acronym}
\usepackage{algorithm}
\usepackage{algpseudocode}
\usepackage[english]{babel}
\let\labelindent\relax
\usepackage[inline]{enumitem}
\usepackage[figurename=Fig., tablename=Tab.]{caption}
\usepackage{multirow}
\usepackage{comment}
\usepackage{booktabs}

\usepackage[T1]{fontenc}
\usepackage{graphicx}
\usepackage{cite}
\usepackage{caption}
\usepackage{subfigure}
\usepackage{epstopdf}
\usepackage{overpic}
\usepackage{amssymb}
\usepackage{amsmath}
\usepackage{amsthm}
\usepackage{array}
\usepackage{color}
\usepackage{url}
\usepackage{comment}

\usepackage{acronym}
\usepackage{algorithm}
\usepackage{algpseudocode}

\usepackage[english]{babel}
\usepackage[inline]{enumitem}

\usepackage{multirow}
\usepackage{widetext}
\usepackage{cuted}
\usepackage{booktabs}
\usepackage{siunitx}

\theoremstyle{plain}

\newtheorem{remark}{Remark}

\newcommand{\figuresname}{Figs.}
\newcommand{\sectionname}{Sect.}
\newcommand{\sectionsname}{Sects.}

\renewcommand{\tablename}{Table}

\newcommand{\mus}{\,\mu\mathrm{s}}
\newcommand{\meter}{\,\mathrm{m}}
\newcommand{\km}{\,\mathrm{km}}
\newcommand{\cm}{\,\mathrm{cm}}

\newcommand{\MHz}{\,\mathrm{MHz}}
\newcommand{\GHz}{\,\mathrm{GHz}}

\newcommand{\MBaud}{\,\mathrm{MBaud}}
\newcommand{\Watt}{\,\mathrm{W}}
\newcommand{\Htran}{\mathsf{H}}
\newcommand{\Ttran}{\mathsf{T}}

\title{Modeling and Performance Evaluation of Distributed Satellite Antennas: \\ Space Is Much Closer Than One Might Think
}

\author{Giacomo~Bacci,~\IEEEmembership{Senior Member,~IEEE,} Riccardo~De~Gaudenzi,~\IEEEmembership{Life~Fellow,~IEEE,}
         Marco~Luise,~\IEEEmembership{Life~Fellow,~IEEE,} and~Luca~Sanguinetti,~\IEEEmembership{Fellow,~IEEE} 
\thanks{The authors are with the University of Pisa, Dipartimento di Ingegneria dell’Informazione, Via Caruso 16, 56122 Pisa, Italy. This
work was supported
in part by the Italian Ministry of Education and Research (MUR) in the
Framework of the FoReLab Project (Departments of Excellence).”}}

\begin{document}
\maketitle

\acrodef{3GPP}{third generation partnership project}
\acrodef{6G}{sixth generation of cellular networks}
\acrodef{5G}{fifth generation of cellular networks}
\acrodef{5-6G}{fifth and sixth generation of cellular networks}
\acrodef{AWGN}{additive white Gaussian noise}
\acrodef{B5G}{beyond-5G}
\acrodef{CF-MIMO}{cell-free massive \acs{MIMO}}
\acrodef{DPD} {disk Poisson Process} 
\acrodef{DRA}{direct radiating antenna}
\acrodef{DC}{direct current}
\acrodef{eMBB}{enhanced mobile broadband}
\acrodef{E-GPW}{extended gridded population of world database}
\acrodef{FDM}{frequency division multiplexing}
\acrodef{FIR}{finite impulse response}
\acrodef{TFoA}{thinned \acs{FoA}}
\acrodef{FoA}{formation of arrays}
\acrodef{GEO}{geostationary Earth orbit}
\acrodef{GSO}{geosynchronous Earth orbit}
\acrodef{H-RRM}{heuristic-\acs{RRM}}
\acrodef{KPI}{key performance indicator}
\acrodef{IF}{intermediate frequency}
\acrodef{LEO}{low Earth orbit}
\acrodef{LoS}{line-of-sight}
\acrodef{MAI}{multiple access interference}
\acrodef{MB}{multi beam}
\acrodef{MEO}{medium Earth orbit}
\acrodef{mMTC}{massive machine-type communications}
\acrodef{eMTC}{enhanced machine-type communications}
\acrodef{PDD}{Poisson disk distribution}
\acrodef{EIRP}{effective isotropic radiated power}
\acrodef{NTN}{non terrestrial network}
\acrodef{URLLC}{ultra-reliable low-latency communications}
\acrodef{ECEF}{Earth-centered Earth-fixed}
\acrodef{GPS}{global positioning system}
\acrodef{HTFS}{high throughput fractionated satellite}
\acrodef{UHF}{ultra-high frequency}
\acrodef{FF}{far-field}
\acrodef{NF}{near-field}
\acrodef{MIMO}{multiple-input multiple-output}
\acrodef{mMIMO}{massive \acs{MIMO}}
\acrodef{GNC}{guidance navigation and control}
\acrodef{GNSS}{global navigation satellite system}
\acrodef{OFDM}{orthogonal frequency division multiplexing}
\acrodef{PHY}{physical-layer}
\acrodef{SA}{sub-array}
\acrodef{SADM}{solar array drive mechanism}
\acrodef{SG}{solar generator}
\acrodef{SNR}{signal-to-noise ratio}
\acrodef{SIR}{signal-to-interference ratio}
\acrodef{SINR}{signal-to-interference-plus-noise ratio}
\acrodef{SSPA}{solid-state power amplifier}
\acrodef{RF}{radio frequency}
\acrodef{R-GEO}{regional \acs{GEO}}
\acrodef{UT}{user terminal}
\acrodef{UE}{user equipment}
\acrodef{QVD}{quantized virtual distancing}
\acrodef{R-GEO}{regional \acs{GEO}}
\acrodef{RRM}{radio resource management}
\acrodef{SSPA}{solid-state power amplifier}
\acrodef{BLER}{block error rate}
\acrodef{DL}{downlink}
\acrodef{UL}{uplink}
\acrodef{PSD}{power spectral density}
\acrodef{UDSM}{ultra deep sub-micron}
\acrodef{PEPS}{power energy platform simulation}
\acrodef{SE}{spectral efficiency}
\acrodef{wrt}{with respect to}
\acrodef{OBP}{on-board digital processor}
\acrodef{UDSM}{ultra-deep-sub-micron}
\acrodef{3GPP}{third generation partnership project}
\acrodef{AWGN}{additive white Gaussian noise}
\acrodef{B5G}{beyond-5G}
\acrodef{CS}{central satellite}
\acrodef{DRA}{direct radiating antenna}
\acrodef{DC}{direct current}
\acrodef{DoD}{direction of departure}
\acrodef{eMBB}{enhanced mobile broadband}
\acrodef{FDM}{frequency division multiplexing}
\acrodef{FIR}{finite impulse response}
\acrodef{TFoA}{thinned \acs{FoA}}
\acrodef{FoA}{formation of arrays}
\acrodef{GEO}{geostationary Earth orbit}
\acrodef{GSO}{geosynchronous Earth orbit}
\acrodef{KPI}{key performance indicator}
\acrodef{IF}{intermediate frequency}
\acrodef{LEO}{low Earth orbit}
\acrodef{LoS}{line-of-sight}
\acrodef{MAI}{multiple access interference}
\acrodef{MEO}{medium Earth orbit}
\acrodef{mMTC}{massive machine-type communications}
\acrodef{eMTC}{enhanced machine-type communications}
\acrodef{EIRP}{effective isotropic radiated power}
\acrodef{NTN}{non terrestrial network}
\acrodef{URLLC}{ultra-reliable low-latency communications}
\acrodef{ECEF}{Earth-centered Earth-fixed}
\acrodef{GPS}{global positioning system}
\acrodef{HTFS}{high throughput fractionated satellite}
\acrodef{HTS}{high throughput satellite}
\acrodef{UHF}{ultra-high frequency}
\acrodef{MIMO}{multiple input multiple output}
\acrodef{GNC}{guidance navigation and control}
\acrodef{NTN}{nonterrestrial network}
\acrodef{GNSS}{global navigation satellite system}
\acrodef{OFDM}{orthogonal frequency division multiplexing}
\acrodef{PHY}{physical-layer}
\acrodef{SADM}{solar array drive mechanism}
\acrodef{SG}{solar generator}
\acrodef{SNR}{signal-to-noise ratio}
\acrodef{SIR}{signal-to-interference ratio}
\acrodef{SINR}{signal-to-interference-plus-noise ratio}
\acrodef{SSPA}{solid-state power amplifier}
\acrodef{RF}{radio frequency}
\acrodef{R-GEO}{regional \acs{GEO}}
\acrodef{UT}{user terminal}
\acrodef{UE}{user equipment}
\acrodef{TDM}{time division multiplexing}
\acrodef{RRM}{radio resource management}
\acrodef{SSPA}{solid-state power amplifier}
\acrodef{BLER}{block error rate}
\acrodef{DL}{down-link}
\acrodef{UL}{up-link}
\acrodef{RV}{random variable}
\acrodef{PSD}{power spectral density}
\acrodef{UDSM}{ultra deep sub-micron}
\acrodef{PEPS}{power energy platform simulation}
\acrodef{SE}{spectral efficiency}
\acrodef{wrt}{with respect to}
\acrodef{OBP}{on-board digital processor}
\acrodef{UDSM}{ultra-deep-sub-micron}
\acrodef{UC}{user-centric}
\acrodef{CSI}{channel state information}
\acrodef{PAC}{per-antenna constraint}
\acrodef{FPAC}{fair per-antenna constraint}
\acrodef{MPC}{maximum power constraint}
\acrodef{AWGN}{additive white Gaussian noise}
\acrodef{TX}{transmit}
\acrodef{MMSE}{minimum mean square error}
\acrodef{MF}{matched filter}
\acrodef{ZF}{zero forcing}
\acrodef{ELSA}{enhanced logarithmic spiral array}
\acrodef{UPA}{uniform planar array}
\acrodef{NB}{narrowband}
\acrodef{WB}{wideband}
\acrodef{BFN}{beamforming network}
\acrodef{MD-MIQP}{minimum distance mixed integer quadratic problem}
\acrodef{PDF}{probability density function}
\acrodef{NPR}{noise-to-power ratio}
\acrodef{HPA}{high power amplifier}
\acrodef{CF}{cell-free}
\acrodef{3GPP}{third generation partnership project}
\acrodef{UC-MIMO}{user-centric \acs{MIMO}}
\acrodef{VHTS}{very high throughput satellites}
\acrodef{PM-MIMO}{pragmatic \acs{mMIMO}}
\acrodef{NR}{new radio}
\acrodef{LMS}{land mobile satellite}
\acrodef{IM}{intermodulation}
\acrodef{OBO}{output backoff}
\acrodef{FDD}{frequency division duplexing}
\acrodef{TDD}{time division duplexing}
\acrodef{BLER}{block error rate}
\acrodef{rv}{random variable}
\acrodef{COB}{center-of-beam}
\acrodef{ISL}{inter-satellite link}
\acrodef{AP}{access point}
\acrodef{SVD}{single value decomposition}
\acrodef{CW}{continuous wave}
\acrodef{AOCS}{attitude and orbit control system}
\acrodef{iid}[i.i.d.]{independent and identically distributed}
\acrodef{OISL}{optical \acs{ISL}}
\acrodef{PAT}{pointing, acquisition, and tracking}
\acrodef{SWaP}{size, weight, and power}
\acrodef{Tbps}{terabit per second}
\acrodef{RTD}{round-trip-delay}
\acrodef{OFDM}{orthogonal frequency division multiplexing}
\acrodef{AI}{artificial intelligence}
\acrodef{TTD}{true time delay}
\acrodef{LO}{local oscillator}
\acrodef{D2D}{direct-to-device}
\acrodef{GW}{gateway}
\acrodef{BH}{beam-hopping}
\acrodef{MSS}{mobile-satellite-services}
\acrodef{QoS}{quality of service}
\acrodef{FFT}{fast Fourier transform}
\acrodef{SM}{spatial multiplexing}
\acrodef{PA}{phased-array}
\acrodef{QVD}{quasi-virtual distancing}
\acrodef{3D}{three-dimensional}
\acrodef{2D}{bidimensional}
\acrodef{sFF}{stacked \acs{FF}}
\begin{abstract}

Distributed satellites can deliver broadband connectivity directly to commercial handsets. In the \ac{FoA} architecture, cooperating satellites synthesize a large aperture whose extent depends on their number, arrangement, and spacing, with intersatellite distances ranging from tens of meters to hundreds of kilometers. For representative S-band geostationary and low-Earth-orbit configurations, the Fraunhofer distance may exceed the nadir link range at spacings on the order of a hundred meters. Therefore, accurate assessment of \ac{FoA} performance across satellite populations and spacings requires spherical-wave modeling for users in the radiative near field, below the formation's Fraunhofer distance. Planar models of satellite positions, array orientations, and user locations must also be assessed against spherical geometries accounting for orbital and Earth curvature. This paper combines these models to quantify how simplifying assumptions affect throughput and outage probability. Numerical results show that multi-antenna precoding based on a local plane-wave approximation within each satellite array closely matches the spherical-wave reference. A global plane-wave approximation can instead cause severe losses, including full outage. In the investigated scenarios, formation curvature has little impact, whereas neglecting Earth curvature overestimates low-Earth-orbit throughput. These findings clarify when geometric and propagation approximations remain reliable and demonstrate that, for large distributed antenna apertures, \emph{space is much closer than one might think}.



\begin{IEEEkeywords}
Direct-to-device communications, 
formation of arrays, Fraunhofer distance, near-field communications,
planar and spherical space- and user-segment modeling, precoding.
\end{IEEEkeywords}

\end{abstract}

\section{Introduction}
Integrating satellite and terrestrial networks enables ubiquitous and resilient connectivity \cite{RdG_2022,Hossein_2021,RdG_2026}. In particular, \ac{D2D} \ac{NTN} services linking unmodified commercial handsets directly to satellites have reached commercial deployment \cite{Laursen_2023}. Delivering broadband throughput to these power- and antenna-limited terminals requires high satellite antenna gain and many narrow beams to reuse the scarce \acl{MSS} spectrum spatially. Both requirements call for a very large satellite aperture.

Launcher-fairing and deployment constraints limit the size of monolithic on-board antennas. A compelling alternative is to distribute the aperture across cooperating satellites, each carrying a moderate-size array \cite{Bekey_2006,Deng_2021,Nanzer_2021}. In this \acf{FoA} paradigm \cite{Bacci_2023,Bacci_2024,RdG_2025}, the number, arrangement, and spacing of satellites determine the distributed aperture, whose spatial extent governs angular resolution and the potential for frequency reuse.

\subsection{Motivation and Contribution}

Assessing the potential of a \ac{FoA} requires addressing several challenges, including synchronization and formation control \cite{RdG_2026,RdG_WSA2026}, wideband beamforming \cite{Bacci_2024}, \acl{CSI} acquisition \cite{RdG_2025,Guidotti_2024}, and coverage constraints imposed by a limited number of simultaneous narrow beams \cite{RdG_WSA2026,RdG_2026}. Beyond these and other implementation challenges, a fundamental modeling question cannot be overlooked: whether ground terminals lie in the \ac{FF} or the radiative \ac{NF} of the FoA \cite{Balanis2016}.

The large propagation distances of satellite links naturally suggest that a plane-wave approximation should suffice. However, unlike conventional satellite antennas, a \ac{FoA} can span hundreds of meters or kilometers \cite{Bacci_2023,Bacci_2024,RdG_2025}. The propagation regime depends on the link distance relative to the \emph{Fraunhofer distance} of the entire aperture \cite{Balanis2016}, which scales quadratically with its extent and inversely with the wavelength. This extent depends on the number of satellites, their arrangement, and the intersatellite spacing. Larger formations and higher carrier frequencies can therefore place ground terminals in the radiative \ac{NF}, despite their considerable distance from the satellites.

For the \ac{GEO} and \ac{LEO} systems considered in \cite{RdG_2025} and operating at \(f_0=2.2\GHz\), the nadir \ac{NF}-\ac{FF} transition occurs at intersatellite spacings of only about one hundred meters; see \sectionname~\ref{channel:fraunhofer}. These values are well below the kilometer-scale and larger spacings considered in the literature, including our own work \cite{RdG_2025}. At higher carrier frequencies, this transition occurs at even smaller spacings. Thus, \emph{space is much closer than one might think}: neglecting wavefront curvature can distort beam patterns and compromise multi-antenna precoding, leading to misleading performance assessments.

Accurate channel modeling also requires consistent space- and user-segment geometries. A planar formation model can alter satellite positions and array orientations, affecting propagation phases and element gains. Likewise, neglecting Earth curvature, particularly in \ac{LEO}, can distort coverage, user distribution, and propagation distances. Geometric approximations must therefore be assessed alongside the propagation model to establish which simplifications preserve reliable performance predictions.

Building on \cite{Bacci_2023,Bacci_2024,RdG_2025}, this paper uses spherical geometries and exact spherical-wave propagation as a reference to assess system performance in terms of throughput and outage probability. Its main contributions are:
\begin{itemize}

\item \emph{Consistent space- and user-segment geometry.}
We develop spherical models of satellite positions, array orientations, and user deployment, and assess their planar approximations. In the investigated scenarios, formation curvature has little impact on throughput, whereas neglecting Earth curvature overestimates \ac{LEO} throughput.

\item \emph{Closed-form characterization of the \ac{NF}-\ac{FF} boundary.}
We derive closed-form Fraunhofer-distance approximations for conformal formations, recover the planar expressions, and determine the intersatellite spacing at which a nadir user enters the radiative near-field.

\item \emph{Performance impact of channel-model mismatch.}
We compare precoders based on exact spherical-wave propagation, local plane-wave approximations within individual satellite arrays, and a global plane-wave approximation across the formation, evaluating all three over the same exact channel. Beam-pattern analysis and Monte Carlo simulations show that locally approximated precoding closely matches the exact-model benchmark when users lie in each array's \ac{FF} and exact satellite-center-to-user distances are retained. In contrast, precoding based on the global approximation can severely reduce throughput and increase outage probability, reaching full outage in the investigated configurations.

\end{itemize}

\subsection{Related Work and Positioning}
\label{intro:sota}

Distributed satellite antennas have been widely investigated to improve coverage and capacity \cite{Bakhsh_2024,Marrero_2022}. Existing studies provide key building blocks for \ac{FoA} systems, including formation architectures, propagation models, and cooperative transmission schemes. Our focus is their integration: assessing a large distributed aperture requires consistent modeling of satellite positions, array orientations, user deployment, and propagation.

Existing channel models capture different aspects of this geometry. Distance-based \acl{LoS} models have been used for capacity and single-user beamforming analyses \cite{Deng_2021,Xu_2024}, while scalar Rician models support network-level evaluations without explicitly resolving the distributed antenna response \cite{Abdelsadek_2021,Abdelsadek_2022,Abdelsadek_2023,Palou_2022,Omid_2023}. Other approaches stack the local \ac{FF} responses of individual satellite arrays \cite{Roeper_2022,Guidotti_2024,Wang_2025,ZhangKim2025,RamezaniBjornson2026}. When geometric phase differences between satellites are retained and the user lies in the \ac{FF} of each array, this construction can accurately approximate propagation even within the \ac{NF} of the overall formation. We refer to this approximation as the \acf{sFF} model. Suitable channel representations therefore already exist; their use must nevertheless be assessed together with the underlying geometric assumptions.

Global plane-wave models are also used in satellite-swarm and \ac{FoA} studies \cite{Tuzi_2023,Beidas_2026,Artiga_2026,Bacci_2023,Bacci_2024,RdG_2025}. Their applicability depends on the formation aperture, as also discussed in \cite{SatNex_YB}. Our previous work \cite{RdG_2025}, for example, considered spacings beyond the \ac{FF} validity range identified here. Designing and evaluating a precoder using the same approximate channel does not expose its mismatch with spherical-wave propagation. A consistent assessment must therefore examine both the geometry used to construct the channel and the propagation model used to evaluate transmission.

The electromagnetic foundations of near-field propagation are well established \cite{Sherman_1962,Balanis2016}, and its implications for large-array communications have been investigated in terrestrial settings, including channel-model mismatch \cite{Bacci_2023wcl}. Related work has also considered distributed ground-station reception from satellites \cite{Vennam2025}. These studies, however, do not directly address the joint geometric and propagation assumptions underlying a distributed transmitting aperture in orbit. The issue is not the absence of suitable channel models: distance-based and appropriately constructed local \ac{FF} models can already capture propagation across such an aperture. Rather, it is the need to assess them jointly with the space- and user-segment geometries. Our unified framework makes these assumptions explicit, examines their validity limits, and evaluates beamforming against a common spherical-wave reference, distinguishing the effects of geometric simplification from propagation-model mismatch on communication-system \acp{KPI}.



\subsection{Reproducible Research}
\label{intro:reproducibility}

We provide freely downloadable MATLAB code\footnote{The link to the GitHub repository will be made available upon completion of the review process.} for the space and user geometries, the three channel models, and the complete simulation framework for evaluating beam patterns, throughput, and outage probability. The release includes the configurations needed to reproduce the reported results and explore alternative system parameters.

\subsection{Organization and Notation}
\label{intro:organization}

The remainder of the paper is organized as follows. \sectionsname~\ref{model:spaceSegment} and \ref{model:userSegment} present the space- and user-segment geometries, respectively. \sectionname~\ref{sec:channel} characterizes the Fraunhofer distance, presents the three channel models, and compares their beam patterns in \sectionname~\ref{results:beamPattern}. \sectionname~\ref{sec:results} evaluates system performance, and \sectionname~\ref{sec:conclusion} concludes the paper. The main derivations are provided in the appendices. 

Boldface lowercase and uppercase letters denote column vectors and matrices, respectively. \(\mathbf{I}_N\) is the \(N\times N\) identity matrix; \((\cdot)^\Ttran\) and \((\cdot)^\Htran\) denote transpose and conjugate transpose; \(\|\cdot\|_2\) is the Euclidean norm. Finally, \(\lfloor\cdot\rfloor\) and \(\operatorname{mod}(\cdot,\cdot)\) denote the floor and modulo operations.

\section{Space-Segment Modeling}
\label{model:spaceSegment}
\begin{figure}[t]
\centering
\includegraphics[width=\columnwidth]{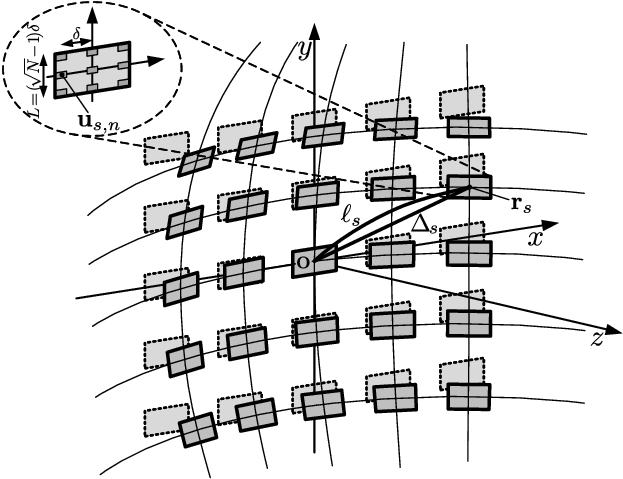}
\caption{Conformal formation with $S=25$ satellites, each hosting a \acs{UPA} with $N=9$ elements, at orbital radius $R$.}
\label{fig:conformal}
\end{figure}

Let us consider a formation of $S$ satellites orbiting at radius
$R=R_\mathsf{E}+h_\mathsf{sat}$, where $R_\mathsf{E}=6{,}371\km$ is the Earth radius and $h_\mathsf{sat}$ is the orbital altitude. As illustrated in \figurename~\ref{fig:conformal}, the formation center defines the origin $\mathbf{O}=[0,0,0]^\Ttran$, while the Earth center is located at $\mathbf{C}=[0,0,R]^\Ttran$. Each satellite carries a square \ac{UPA} with $N$ uniformly spaced elements, arranged in $\sqrt{N}$ rows by $\sqrt{N}$ columns with spacing $\delta$. The resulting array side length is
$L=(\sqrt{N}-1)\delta$, constrained by the launcher-fairing dimensions. The position of the $n$th element on satellite $s$, whose center is $\mathbf{r}_s$, is denoted by $\mathbf{u}_{s,n}\in\mathbb{R}^3$, for $n=1,\ldots,N$ and $s=1,\ldots,S$.

Although a regular \emph{planar} $\sqrt{S}\times\sqrt{S}$ grid is conceptually convenient, maintaining it in orbit would require continuous station keeping and, over a finite region, its geometry cannot be mapped onto a sphere without distortion \cite{Oneill_2006}. Next, we develop a conformal \emph{spherical} placement.

\subsection{Spherical Formation Geometry}
\label{spaceSegment:conformal}

We adopt a \emph{radially equidistant} construction. Let
\begin{align}\label{eq:phi}
  \varphi=2\sin^{-1}\left(\frac{\Delta}{2R}\right)
\end{align}
be the angle subtended at the Earth center by two adjacent satellites of the central row (or, equivalently, of the central column) separated by a chord of length $\Delta$, which corresponds to a geodesic separation $R\varphi$. For the generic satellite $s$, we then require the geodesic distance from the \ac{FoA} center (i.e., the \emph{arc} between $\mathbf{r}_s$ and $\mathbf{O}$) to be
\begin{align}\label{eq:ell_s}
  \ell_s=R\varphi \sqrt{i_s^2+j_s^2}
\end{align}
where
\begin{align}
\label{eq:i}
i_s&=\textrm{mod}\left(s-1,\sqrt{S}\right)-\frac{\sqrt{S}-1}{2}\\
\label{eq:j}
j_s&=\left\lfloor\frac{s-1}{\sqrt{S}}\right\rfloor-\frac{\sqrt{S}-1}{2}
\end{align}
are the horizontal and vertical indices of satellite $s$ in the reference grid. As derived in \appendixname~\ref{app:conformal_satellite}, the position of the $s$th satellite array center is $\mathbf{r}_s=[x_s, y_s, z_s]^\Ttran$, with
\begin{align}
\label{eq:conformal_satellite_x}
x_s&=\hat{x}_s\cdot R\sin\theta_s\\
\label{eq:conformal_satellite_y}
y_s&=\hat{y}_s\cdot R\sin\theta_s\\ 
\label{eq:conformal_satellite_z}
z_s&=R\left[1-\cos\theta_s\right]
\end{align}
where $\theta_s=\ell_s/R=\varphi \sqrt{i_s^2+j_s^2}$, and
\begin{align}
\label{eq:hat}
(\hat{x}_s,\hat{y}_s)
=\frac{(i_s,j_s)}{\sqrt{i_s^2+j_s^2}},
\qquad (i_s,j_s)\neq(0,0).
\end{align}
For the central satellite, $\mathbf{r}_s=\mathbf{O}$. The resulting spherical geometry is illustrated in \figurename~\ref{fig:conformal}, which also reports the distances $\ell_s$ and the corresponding chords $\Delta_s=\sqrt{i_s^2+j_s^2}\cdot \Delta$; the curved grid lines are included only for visualization and, except for the central row and column, are not geodesics.

\begin{figure}[t]
\centering
\subfigure[]
{\includegraphics[width=0.45\columnwidth]
{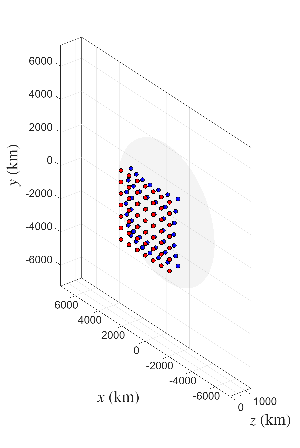}
\label{fig:model_conf_vs_planar_a}}
\subfigure[]
{\includegraphics[width=0.45\columnwidth]
{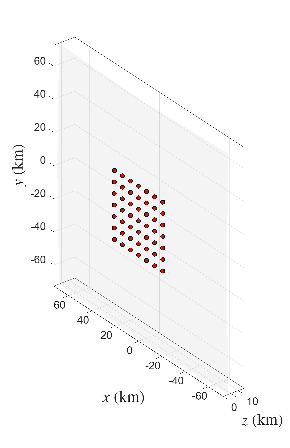}
\label{fig:model_conf_vs_planar_b}}
\caption{Planar (red) and conformal (blue) configurations for $S=49$ \ac{LEO} satellites at $h_\mathsf{sat}=550\km$: \subref{fig:model_conf_vs_planar_a} $\Delta=R/10$ and \subref{fig:model_conf_vs_planar_b} $\Delta=R/1000$.}
\label{fig:model_conf_vs_planar}
\end{figure}

\begin{remark}[Planar configuration]
\label{rem:planar}
When $\max_s\theta_s\ll1$, we have $\varphi\simeq\Delta/R$, and \eqref{eq:conformal_satellite_x}-\eqref{eq:conformal_satellite_z} reduce to
\begin{align}
\mathbf{r}_s\simeq[i_s\Delta,j_s\Delta,0]^\Ttran
\end{align}
recovering the regular planar grid commonly adopted in the literature \cite[Eqs.~(4)-(6)]{Bacci_2023}. The corresponding planar satellites are shown as dashed light-gray boxes in \figurename~\ref{fig:conformal}.

As an example, \figurename~\ref{fig:model_conf_vs_planar} compares the planar and conformal configurations for a \ac{LEO} formation with $h_\mathsf{sat}=550\km$ and $S=49$. For $\Delta=R/1000=6.92\km$, their maximum separation is only $62.3\meter$, whereas for the illustrative large spacing $\Delta=R/10=692.1\km$, it reaches $620.1\km$.
\end{remark}

\subsection{Satellite-Array Geometry}
\label{spaceSegment:array}

Having specified the satellite-center positions $\mathbf{r}_s$, we now determine the positions of the $N$ antenna elements carried by each satellite. Since the array size satisfies $L\ll R$, the curvature across an individual array is negligible. We therefore model each \ac{UPA} as a planar array tangent to the orbital sphere at $\mathbf{r}_s$. As derived in \appendixname~\ref{app:conformal_antenna}, the position of element $n$ on satellite $s$ is
\begin{align}
\label{eq:conformal_antenna}
\mathbf{u}_{s,n}=\mathbf{r}_s+\mathbf{P}_s\mathbf{r}_n
\end{align}
where $\mathbf{r}_n=[i_n\delta,j_n\delta,0]^\Ttran$ and
\begin{align}
\label{eq:i_n}
i_n&=\operatorname{mod}(n-1,\sqrt{N})-\frac{\sqrt{N}-1}{2}\\
j_n&=\left\lfloor\frac{n-1}{\sqrt{N}}\right\rfloor-\frac{\sqrt{N}-1}{2}.
\end{align}
The rotation matrix
\begin{align}
\label{eq:rotation_matrix}
\mathbf{P}_s=
\left[\mathbf{a}_s\quad\boldsymbol{\mu}_s\quad\boldsymbol{\nu}_s\right]
\end{align}
orients the array along the local azimuthal, meridional, and inward-normal directions, computed as
\begin{align}
\label{eq:azimuthal_direction}
\mathbf{a}_s&=[-\hat{y}_s,\hat{x}_s,0]^\Ttran\\
\label{eq:meridional_direction}
\boldsymbol{\mu}_s&=[-\hat{x}_s\cos\theta_s,-\hat{y}_s\cos\theta_s,
-\sin\theta_s]^\Ttran\\
\label{eq:normal_direction}
\boldsymbol{\nu}_s&=[-\hat{x}_s\sin\theta_s,-\hat{y}_s\sin\theta_s,
\cos\theta_s]^\Ttran.
\end{align}
The tangent-plane constraint fixes the array boresight $\boldsymbol{\nu}_s$ but leaves one rotational degree of freedom about it, which we remove by aligning the local array axes with $\mathbf{a}_s$ and $\boldsymbol{\mu}_s$.

For the central satellite, we set $\mathbf{P}_s=\mathbf{I}_3$. The planar configuration is obtained by setting
$\mathbf{r}_s=[i_s\Delta,j_s\Delta,0]^\Ttran$
and aligning all arrays with the global axes,
i.e., $\mathbf{P}_s=\mathbf{I}_3$.

\subsection{Formation and Constellation Architecture}
\label{spaceSegment:architecture}
The spherical construction above supports both fully and partially populated reference grids. In a fully populated formation, all nodes of the $\sqrt{S}\times\sqrt{S}$ grid are occupied, with indices defined by \eqref{eq:i}. In a partially populated formation, only a subset of the reference-grid nodes is occupied, while the position of each selected satellite is still given by \eqref{eq:conformal_satellite_x}-\eqref{eq:conformal_satellite_z}. This provides greater flexibility in trading the number of satellites against the synthesized aperture. The \ac{LEO} reference system considered in \cite{RdG_2025}, for example, is obtained by removing the four corners of a $3\times3$ grid with pitch $\Delta$. The $S=5$ remaining satellites occupy $
(i_s,j_s)\in\{(0,0),(\pm1,0),(0,\pm1)\} $ forming a \emph{cross} whose four outer array centers are at distance $\Delta$ from the formation center. 

The proposed model describes the local geometry of a single formation. At the constellation level, multiple formations may be deployed, for example according to a Walker-Delta pattern \cite{Richharia_2010}, with each formation acting as a large-aperture virtual satellite. The resulting hierarchical architecture is characterized by an intra-formation spacing $\Delta_1=\Delta$, and a much larger inter-formation spacing $\Delta_2\gg\Delta_1$. Although we focus on curvature effects within each formation, the same modeling approach applies across formations by accounting for their relative positions and array orientations.

\section{User-Segment Modeling}\label{model:userSegment}
\begin{figure}[t]
\centering
\includegraphics[width=\columnwidth]{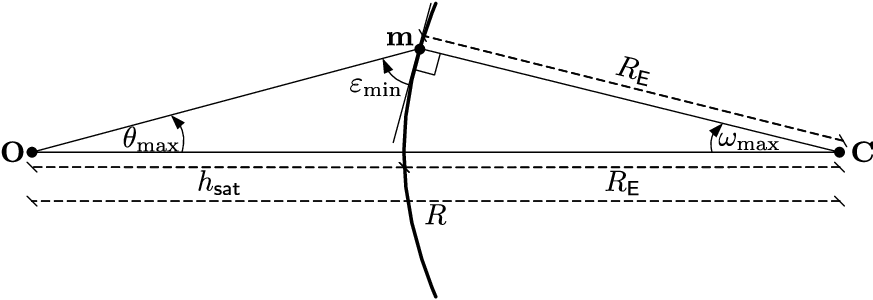}
\caption{Coverage geometry for an edge \ac{UT} at $\mathbf{m}$ \cite{Richharia_2010}.}
\label{fig:coverage}
\end{figure}

We consider $K$ single-antenna \acp{UT}, jointly served by the satellite system over the same time-frequency resource. The \acp{UT} are randomly distributed within a coverage area centered at the nadir point of the central satellite. Coverage requires an elevation angle of at least $\varepsilon_{\rm min}$, which determines the maximum nadir angle
\begin{align}
\label{eq:nadir_angle}
\theta_{\rm max}
=\sin^{-1}\left(\frac{R_\mathsf{E}}{R}\cos\varepsilon_{\rm min}\right)
\end{align}
and the corresponding maximum Earth-central angle
\begin{align}
\label{eq:central_angle}
\omega_{\rm max}
=\frac{\pi}{2}-\varepsilon_{\rm min}-\theta_{\rm max}
\end{align}
as illustrated in \figurename~\ref{fig:coverage} \cite{Richharia_2010}. Here, $\theta_{\rm max}$ is measured between the satellite boresight $\hat{\mathbf{z}}=[0,0,1]^\Ttran$ and the line of sight to an edge \ac{UT}, whereas $\omega_{\rm max}$ is measured at the Earth center. For each \ac{UT} $k$, we introduce two independent random variables $U_k,V_k\sim\mathcal{U}[0,1)$. As for the space segment, the spherical geometry is adopted as the reference model, with the flat-Earth model recovered as a limiting case.

The \acp{UT} are placed on the surface of the spherical Earth, with radius $R_\mathsf{E}$ and center $\mathbf{C}=[0,0,R]^\Ttran$, within the spherical cap of angular radius $\omega_{\rm max}$ defined in \eqref{eq:central_angle}. To ensure a uniform density per unit area, \ac{UT} $k$'s position is drawn as
%
\begin{align}
&\mathbf{m}_k=\nonumber\\
&\,\,\left[R_\mathsf{E}\sin\omega_k\cos(2\pi V_k), R_\mathsf{E}\sin\omega_k\sin(2\pi V_k), R\!-\!R_\mathsf{E}\cos\omega_k\right]^\Ttran
\end{align}
and its Earth-central angle is generated as
\begin{align}
\label{eq:cap_angle}
\omega_k
=\cos^{-1}\left[1-U_k\left(1-\cos\omega_{\rm max}\right)\right].
\end{align}
The corresponding coverage radius, measured as a geodesic on the Earth surface, is $r_\mathsf{cap}=R_\mathsf{E}\,\omega_{\rm max}$, whereas its chordal projection on the $x$-$y$ plane is $R_\mathsf{E}\sin\omega_{\rm max}$. Consistently with the spherical geometry, the minimum-distance constraint between \acp{UT} is enforced, as in \cite{RdG_2025}, through the great-circle (geodesic) distance
\begin{align}\label{eq:cap_distance}
  d_{k,k^\prime}=R_\mathsf{E}\cos^{-1}\!\left(\hat{\mathbf{m}}_k^\Ttran\hat{\mathbf{m}}_{k^\prime}\right)\ge d_{\rm min}
\end{align}
where $\hat{\mathbf{m}}_k=(\mathbf{m}_k-\mathbf{C})/R_\mathsf{E}$ is the radial unit vector pointing from the Earth center to the $k$th \ac{UT}.

\begin{figure}[t]
\centering
\includegraphics[width=\columnwidth]{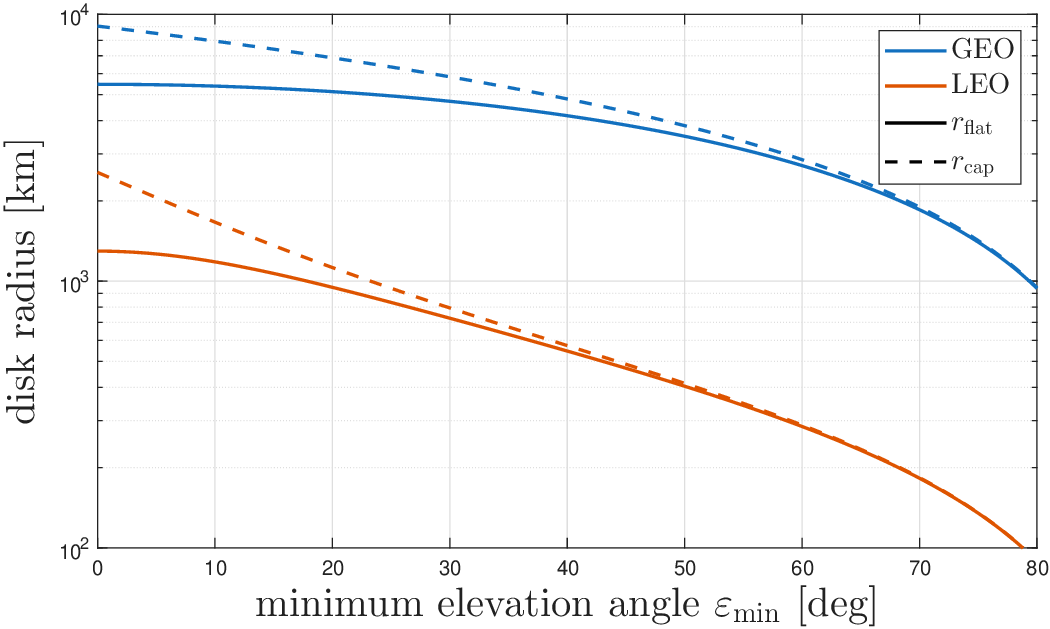}
\caption{Spherical-cap and flat-Earth coverage radii versus the minimum \ac{UT} elevation angle.}
\label{fig:coverage_radius}
\end{figure}

\begin{remark}[Flat-Earth approximation]
\label{rem:flat_user}
When $\omega_{\rm max}\ll1$, the spherical cap is locally approximated by the plane tangent to the Earth at the nadir point, located at $z=h_\mathsf{sat}$. The coverage region becomes a disk of radius
\begin{align}
\label{eq:flat_radius}
r_\mathsf{flat}=h_\mathsf{sat}\tan\theta_{\rm max}
\end{align}
and an area-uniform \ac{UT} position is generated as
\begin{align}
\label{eq:flat_user}
\mathbf{m}_k
&=\left[\rho_k\cos(2\pi V_k),
\rho_k\sin(2\pi V_k),
h_\mathsf{sat}\right]^\Ttran
\end{align}
with $\rho_k =r_\mathsf{flat}\sqrt{U_k}$.
The minimum-distance constraint is then evaluated using
$\|\mathbf{m}_k-\mathbf{m}_{k'}\|_2$. In this limit,
$r_\mathsf{cap}\simeq r_\mathsf{flat}$.

As shown in \figurename~\ref{fig:coverage_radius}, the two models agree at large $\varepsilon_{\rm min}$. For the \ac{LEO} scenario with $\varepsilon_{\rm min}=40^\circ$, however, $r_\mathsf{cap}$ exceeds $r_\mathsf{flat}$ by $26\km$ (about $5\%$). The flat model therefore reduces the coverage of cell-edge \acp{UT} and artificially increases their spatial density, leading to the \ac{KPI} bias discussed in \sectionname~\ref{results:geometry}.
\end{remark}

\section{Channel Modeling}\label{sec:channel}
%
%
%
We now characterize the propagation channel. For a \ac{UT} located at $\mathbf{m}$, let
\begin{align}\label{eq:channel_vector}
  \mathbf{h}(\mathbf{m})=\left[h_{1,1}(\mathbf{m}),\,\dots,\,h_{S,N}(\mathbf{m})\right]^\Ttran\in\mathbb{C}^{M}
\end{align}
be the channel (steering) vector, between the $M=NS$ radiating elements of the satellite formation and the \ac{UT}, with elements
\begin{align}
  \label{eq:channel_coefficient}
  h_{s,n}(\mathbf{m}) =
    \zeta_{s,n}(\mathbf{m})
    \exp\!\left\{\jmath\phi_{s,n}(\mathbf{m})\right\}.
\end{align}
In particular, \eqref{eq:channel_coefficient} corresponds to the channel, relative to  the element $n$ on satellite $s$ located in $\mathbf{u}_{s,n}$, and observed from position $\mathbf{m}$, where $\zeta_{s,n}(\mathbf{m})$ is the gain factor, and $\phi_{s,n}(\mathbf{m})$ is the specific element-to-observer phase shift.

\subsection{Fraunhofer distance}\label{channel:fraunhofer}

Before specifying \eqref{eq:channel_coefficient}, we characterize the \ac{FoA} Fraunhofer distance $z_\mathsf{F}$, beyond which the wavefront is locally planar within a maximum phase error of $\pi/8$. Standard Fraunhofer-distance expressions are typically derived for planar arrays and cannot be directly applied to the conformal geometry considered here. What matters is the \emph{curved} aperture offered by the formation: we therefore rederive $z_\mathsf{F}$ in \appendixname~\ref{app:fraunhofer}, obtaining, for $(\sqrt{S}-1)\Delta\gg L$,
\begin{align}
\label{eq:fraunhofer_conf_main}
z_\mathsf{F}
\approxeq\frac{8R^2}{\lambda}
\sin^2\left[
\sqrt{2}(\sqrt{S}-1)
\sin^{-1}\left(\frac{\Delta}{2R}\right)
\right]
\end{align}
where we recall that $R=R_\mathsf{E}+h_\mathsf{sat}$, with $R_\mathsf{E}=6{,}371\km$ beign the Earth radius and $h_\mathsf{sat}$ being the orbital altitude.
As a consistency check, when
$(\sqrt{S}-1)\Delta/R\ll1$, the curvature becomes negligible and the conformal configuration approaches a planar array. Accordingly, the above expression reduces to
\begin{align}
\label{eq:fraunhofer_plan_main}
\widetilde{z}_\mathsf{F}
=\frac{4}{\lambda}
\left[(\sqrt{S}-1)\Delta\right]^2
\end{align}
which coincides with the standard planar-array result \cite{Balanis2016}.

\begin{figure}[t]
\centering
\includegraphics[width=1.02\columnwidth]{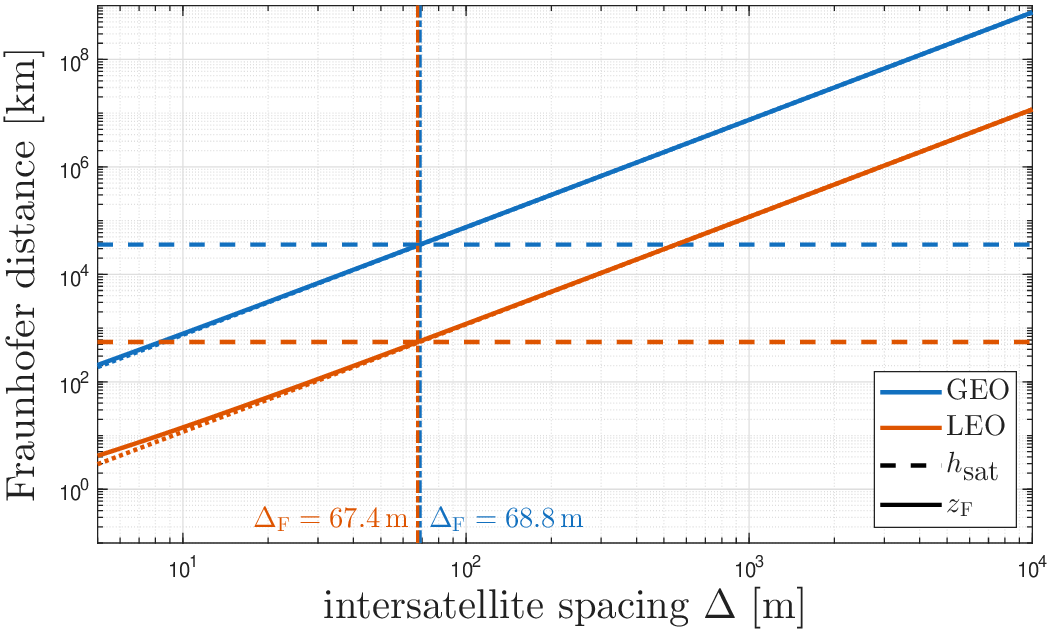}
\caption{Fraunhofer distance versus intersatellite spacing for the reference S-band \ac{GEO} and \ac{LEO} formations. The horizontal dashed lines indicate their orbital altitudes.}
\label{fig:fraunhofer_distance}
\end{figure}

\figurename~\ref{fig:fraunhofer_distance} shows $z_\mathsf{F}$ versus $\Delta$ for the S-band \ac{D2D} systems detailed in \tablename~\ref{tab:GEOvsLEO}, with $\lambda=13.6\cm$: a \ac{GEO} formation with $S=289$ and $h_\mathsf{sat}=35,786\km$, and a \ac{LEO} formation with $S=9$ and $h_\mathsf{sat}=550\km$. These configurations are also used in the system-level evaluation of \sectionname~\ref{sec:results}. For a nadir \ac{UT}, the intersection between $z_\mathsf{F}$ and $h_\mathsf{sat}$ defines the critical spacing $\Delta_\mathsf{F}$: the \ac{FF} approximation applies for $\Delta<\Delta_\mathsf{F}$, whereas \ac{NF} propagation occurs for $\Delta>\Delta_\mathsf{F}$. Using the planar approximation, which is accurate near this transition, we obtain
\begin{align}
\label{eq:DeltaF_square}
\Delta_\mathsf{F}
=\frac{\sqrt{h_\mathsf{sat}\lambda/4}-L}{\sqrt{S}-1}
\end{align}
%
which yields
$\Delta_\mathsf{F}\approxeq68.8\meter$ for the \ac{GEO} system and
$\Delta_\mathsf{F}\approxeq67.4\meter$ for the \ac{LEO} system. Thus, even moderate intersatellite spacings can place terrestrial \acp{UT} in the so-called radiative \ac{NF}. This result confirms one of central messages of this paper: from an electromagnetic perspective, \emph{space is much closer than one might think}. Therefore, the analysis of a \ac{FoA} should start from the exact spherical-wave channel model in the radiative \ac{NF}, with approximate models introduced only after establishing the specific operating conditions under which they remain valid.

\begin{remark}[Fraunhofer Distance for General Formations]
The square reference grid provides a convenient baseline for more general formation geometries. For a nonsquare or irregular arrangement, the Fraunhofer distance can be estimated using an enclosing square aperture. Within the same transverse-aperture criterion, this gives a conservative estimate and avoids deriving a separate expression for every configuration. Configuration-specific expressions can nevertheless provide tighter estimates by accounting for the actual occupied positions, as in the case of the cross-shaped configuration considered in \cite{RdG_2025}.
\end{remark}
\subsection{Radiative near-field model}\label{channel:nf}
When the regime is radiative \ac{NF} (for a nadir observer, this occurs when $h_\mathsf{sat}<z_\mathsf{F}$, i.e., when $\Delta>\Delta_\mathsf{F}$), we have to account for the spherical wavefront across the distributed \ac{FoA} aperture, considering the exact element-to-\ac{UT} distance and the element radiation pattern.
The entry $h_{s,n}(\mathbf{m})$ \eqref{eq:channel_coefficient} of the steering vector, as observed from \ac{UT} position $\mathbf{m}=[x_m, y_m, z_m]^\Ttran$, has the specific element-to-observer phase shift
\begin{align}
  \label{eq:phi_nf}
  \phi_{s,n}(\mathbf{m})=2\pi\,\textrm{mod}\left(\frac{d_{s,n}(\mathbf{m})}{\lambda},1\right)
\end{align}
where
\begin{align}
  \label{eq:d_nf}
  d_{s,n}(\mathbf{m})=\|\mathbf{m}-\mathbf{u}_{s,n}\|
\end{align}
is the (slant) element-to-observer distance, whereas the gain factor in \eqref{eq:channel_coefficient} can be computed as
\begin{align}
  \label{eq:zeta_nf}
  \zeta_{s,n}(\mathbf{m})=\frac{\lambda}{4\pi d_{s,n}(\mathbf{m})}\cdot g\left[\theta_{s,n}(\mathbf{m})\right]
\end{align}
in which the first term is the free-space amplitude attenuation, and the second term is the \acl{DRA} element gain. We assume here that the radiating element is modeled as \cite{Rahmat_1981}
\begin{align}
  \label{eq:g_nf}
  g\left[\theta_{s,n}(\mathbf{m})\right]=\sqrt{\Gamma}\left\{\max\left(0,\cos\left[\theta_{s,n}(\mathbf{m})\right]\right)\right\}^q
\end{align}
where $\Gamma=4\pi\mu A/\lambda^2$ is the \ac{DRA} peak gain, $\mu$ and $A$ denote the \ac{DRA} efficiency and elementary cell area, respectively, $q=(\Gamma-2)/4$, and 
\begin{align}
  \label{eq:theta_nf}
  \theta_{s,n}(\mathbf{m})=\cos^{-1}\left(\hat{\mathbf{m}}^\Ttran_{s,n}\boldsymbol{\nu}_s\right)
\end{align}
is the angle between the local array boresight $\boldsymbol{\nu}_s$, defined in \eqref{eq:normal_direction}, and the \ac{DoD} from radiating element $(s,n)$ towards the \ac{UT}, namely,
\begin{align}
  \label{eq:mhat_nf}
  \hat{\mathbf{m}}_{s,n}=\frac{\mathbf{m}-\mathbf{u}_{s,n}}{d_{s,n}(\mathbf{m})} .
\end{align}

\subsection{Far-field model}\label{channel:ff}

When the propagation occurs in the \ac{FF} regime (for a nadir observer, this occurs when $h_\mathsf{sat}>z_\mathsf{F}$, i.e., when $\Delta<\Delta_\mathsf{F}$), the wavefront across the \ac{FoA} can be locally approximated as planar, so that all radiating elements share the same propagation direction. Moreover, for a sufficiently small angular extent of the \ac{FoA}, i.e., $\max_s\theta_s\ll1$, the local array boresights satisfy $\boldsymbol{\nu}_s\approxeq\hat{\mathbf{z}}=[0,0,1]^\Ttran$ for all $s$. Under these assumptions, the gain factor of each element $h_{s,n}(\mathbf{m})$ \eqref{eq:channel_coefficient} is the same across all radiating elements: 
\begin{align}
  \label{eq:zeta_ff}
  \zeta_{s,n}(\mathbf{m})=\zeta^{\mathsf{FF}}(\mathbf{m})=\frac{\lambda}{4\pi d^\mathsf{FF}(\mathbf{m})}\cdot g\left[\theta^{\mathsf{FF}}(\mathbf{m})\right],
\end{align}
Under the hypothesis of planar wavefront, the free-space amplitude attenuation depends on the distance $d^\mathsf{FF}(\mathbf{m})=\|\mathbf{m}\|$ from the antenna reference point (i.e., the \ac{FoA} center $\mathbf{O}=\left[0,0,0\right]^\Ttran$) to the observer $\mathbf{m}$, and the \ac{DRA} gain \eqref{eq:g_nf}  depends on the angle
\begin{align}
  \label{eq:theta_ff}
  \theta^\mathsf{FF}(\mathbf{m})=\cos^{-1}\left(\hat{\mathbf{m}}^\Ttran\hat{\mathbf{z}}\right)
\end{align}
between the antenna \ac{DoD} $\hat{\mathbf{m}}=\mathbf{m}/d^\mathsf{FF}(\mathbf{m})$ and the global direction $\hat{\mathbf{z}}$, which approximates the local array boresight $\boldsymbol{\nu}_s$ under the small-angular-extent assumption introduced above. Finally, the phase shift can be obtained as
\begin{align}
  \label{eq:phi_ff}
  \phi_{s,n}(\mathbf{m})=\phi^\mathsf{FF}_{s,n}(\mathbf{m})=2\pi\,\textrm{mod}\left(\frac{d^{\mathsf{FF}}(\mathbf{m})}{\lambda},1\right)
  -\frac{2\pi}{\lambda}\mathbf{u}^\Ttran_{s,n}\hat{\mathbf{m}},
\end{align}
where the term $-\frac{2\pi}{\lambda}\mathbf{u}^\Ttran_{s,n}\hat{\mathbf{m}}$ represents the path-length difference of element $(s,n)$ relative to $\mathbf{m}$, based on the Fraunhofer approximation \cite{Balanis2016} $d_{s,n}(\mathbf{m})=\|\mathbf{m}-\mathbf{u}_{s,n}\|=\|\mathbf{m}\|-\mathbf{u}^\Ttran_{s,n}\hat{\mathbf{m}}+\mathcal{O}\left(\frac{\left\|\mathbf{u}_{s,n}\right\|^2}{\|\mathbf{m}\|}\right)$.

\subsection{Stacked far-field model}\label{channel:sff}

When each individual satellite array operates in the \ac{FF}, while the \ac{UT} is still in the \ac{NF} of the overall \ac{FoA}, a convenient intermediate approximation can be introduced. This regime naturally arises when the individual array size is much smaller than the overall \ac{FoA} extent, e.g., when $\Delta\gg L$. We therefore introduce a \emph{piecewise \ac{FF}} model, hereafter referred to as the \ac{sFF} model, in which propagation is modeled exactly across the satellite centers, while a local \ac{FF} approximation is adopted within each satellite array.

To do so, we can modify the relations introduced in \sectionname~\ref{channel:ff}. In particular, the gain factor in \eqref{eq:channel_coefficient},
\begin{align}
  \label{eq:zeta_sff}
  \zeta_{s,n}(\mathbf{m})=\zeta^{\mathsf{sFF}}_s(\mathbf{m})=\frac{\lambda}{4\pi d^\mathsf{sFF}_s(\mathbf{m})}\cdot g\left[\theta^{\mathsf{sFF}}_s(\mathbf{m})\right]
\end{align}
becomes dependent on each satellite $s$. In fact, under the hypothesis of a planar wavefront locally valid at each satellite location $\mathbf{r}_s$, the free-space amplitude attenuation depends on the distance $d^\mathsf{sFF}_s(\mathbf{m})=\|\mathbf{m}-\mathbf{r}_s\|$ from the satellite $s$'s array center to the observer. The \ac{DRA} gain \eqref{eq:g_nf} now depends on the nadir angle
\begin{align}
  \label{eq:theta_sff}
  \theta^\mathsf{sFF}_s(\mathbf{m})=\cos^{-1}\left(\hat{\mathbf{m}}_s^\Ttran\boldsymbol{\nu}_s\right)
\end{align}
between satellite $s$'s \ac{DoD} $\hat{\mathbf{m}}_s=\left(\mathbf{m}-\mathbf{r}_s\right)/d^\mathsf{sFF}_s(\mathbf{m})$ and $\boldsymbol{\nu}_s$. Similarly, the phase shift $\phi_{s,n}(\mathbf{m})$ can be obtained as $  \phi_{s,n}(\mathbf{m})=\phi^\mathsf{sFF}_{s,n}$ with
\begin{align}
  \label{eq:phi_sff}
  \phi^\mathsf{sFF}_{s,n}(\mathbf{m})=2\pi\,\textrm{mod}\left(\frac{d^{\mathsf{sFF}}_s(\mathbf{m})}{\lambda},1\right)
  -\frac{2\pi}{\lambda}\left(\mathbf{u}_{s,n}-\mathbf{r}_s\right)^\Ttran\hat{\mathbf{m}}_s,
\end{align}
where, as in \eqref{eq:phi_ff}, the term $-\frac{2\pi}{\lambda}\left(\mathbf{u}_{s,n}-\mathbf{r}_s\right)^\Ttran\hat{\mathbf{m}}_s$ represents the first-order path-length difference of element $(s,n)$ with respect to the reference path from the center $\mathbf{r}_s$ of satellite $s$ to the observer's location $\mathbf{m}$ (with $d^{\mathsf{sFF}}_s(\mathbf{m})$ already accounting for the path-length between the satellite $s$ and the observer).

\begin{figure}[t]
\centering
\includegraphics[width=0.95\columnwidth]{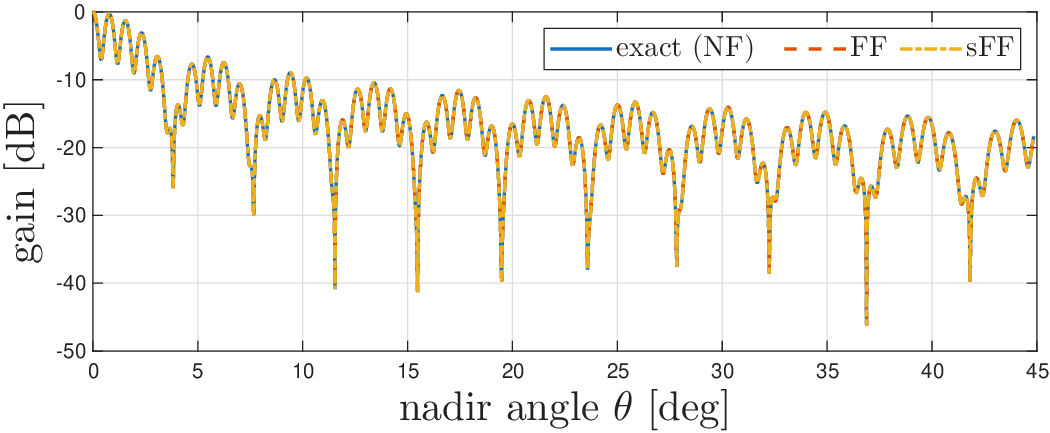}
\caption{\ac{MF} beam patterns for the reference \ac{LEO} formation with $\Delta=10\meter<\Delta_\mathsf{F}$. Precoding uses the \ac{NF}, \ac{FF}, or \ac{sFF} model; propagation is evaluated using the exact \ac{NF} model.}
\label{fig:gain_abs_below}
\end{figure}

\begin{figure}[t]
\centering
\subfigure[]
{\includegraphics[width=0.95\columnwidth]{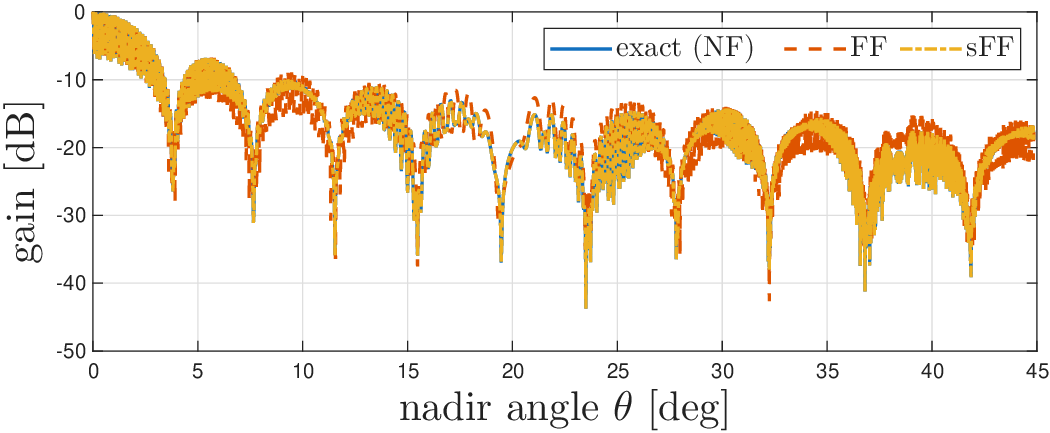}
\label{fig:gain_abs_above}}

\subfigure[]
{\includegraphics[width=0.95\columnwidth]{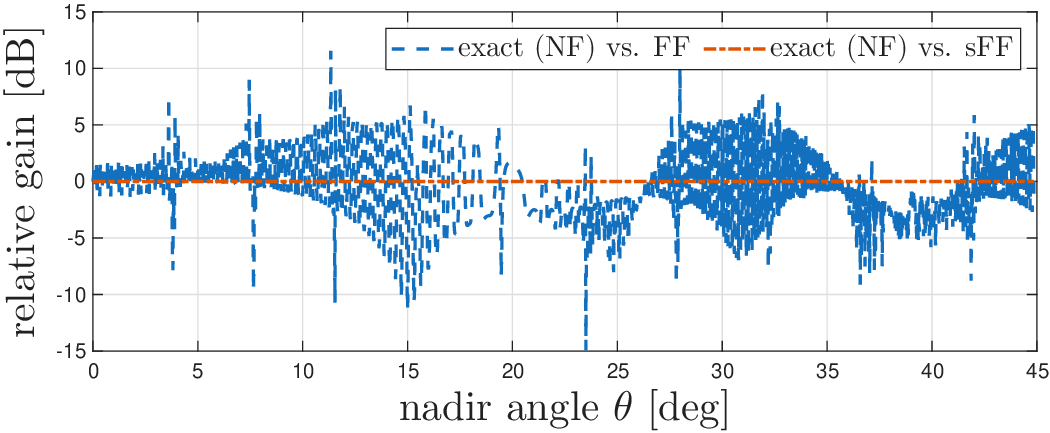}
\label{fig:gain_rel_above}}
\caption{\ac{MF} beam patterns for the reference \ac{LEO} formation with $\Delta=1\km>\Delta_\mathsf{F}$: \subref{fig:gain_abs_above}~beam pattern gains and \subref{fig:gain_rel_above}~gain ratios relative to the \ac{NF}-based reference, for \ac{FF} (dashed) and \ac{sFF} (solid) precoding.}
\label{fig:gain_above}
\end{figure}

\subsection{When Far-Field Beamforming Fails}
\label{results:beamPattern}

We now examine how channel-model mismatch affects the \ac{FoA} beam pattern. We consider the reference \ac{LEO} scenario with a conformal formation of $S=9$ satellites, and evaluate the field on the tangent ground plane described in Remark~\ref{rem:flat_user}. For this scenario, $z_\mathsf{F}=h_\mathsf{sat}$ at $\Delta_\mathsf{F}\approxeq67.4\meter$.
The beam is directed toward the nadir point
$\mathbf{m}_0=[0,0,h_\mathsf{sat}]^\Ttran$, and the pattern is evaluated along the zero-azimuth cut
\begin{align}
\mathbf{m}(\theta)
=[h_\mathsf{sat}\tan\theta,0,h_\mathsf{sat}]^\Ttran.
\end{align}
To compare the channel models, we compute the \ac{MF} precoding vector
$\mathbf{v}(\mathbf{m}_0)=\mathbf{h}'(\mathbf{m}_0)$,
where $\mathbf{h}'(\mathbf{m}_0)$ is generated using the
\ac{NF}, \ac{FF}, or \ac{sFF} model introduced above. The channel
$\mathbf{h}(\theta)=\mathbf{h}(\mathbf{m}(\theta))$ is computed
using the exact \ac{NF} model for every observation angle $\theta$. The \ac{NF}-based precoder is therefore \emph{matched} to the propagation model and serves as the reference, whereas the \ac{FF}- and \ac{sFF}-based precoders are \emph{mismatched}. This comparison isolates the beam-pattern distortion caused by the channel approximations.

We compare $\Delta=10\meter$ and $\Delta=1\km$, respectively below and above the nadir transition spacing
$\Delta_\mathsf{F}\approxeq67.4\meter$.
For $\Delta=10\meter$, the three beam patterns are practically indistinguishable, as shown in \figurename~\ref{fig:gain_abs_below}. For $\Delta=1\km$, however, the \ac{FF}-based pattern deviates substantially from the matched reference, whereas the \ac{sFF}-based pattern remains nearly identical to it; see \figurename~\ref{fig:gain_above}.

In summary, the \ac{sFF} model retains exact satellite-center-to-user distances and applies a plane-wave approximation only within each array, where the \ac{FF} condition holds. It therefore remains accurate in the formation's radiative \ac{NF}, whereas a global plane-wave approximation substantially distorts the beam pattern. Similar results for the \ac{GEO} scenario are omitted for brevity. The next section quantifies the impact of mismatched channel models on throughput and outage probability through system-level simulations.

\section{Numerical evaluation and comparison}\label{sec:results}

We evaluate the impact of modeling assumptions on \ac{FoA}-based \ac{D2D} performance using the S-band scenarios of \cite[\tablename~I]{RdG_2025}, summarized in \tablename~\ref{tab:GEOvsLEO}. To reproduce the results presented here, the MATLAB code freely available offers a full-fledged system-level simulator, which contains the full set of parameters, including \ac{UT} parameters, channel modeling, space- and user-segment configurations, and Monte Carlo generation of \ac{UT} snapshots.



\begin{table}[t!]
    \centering
    \caption{System parameters per scenario.}
    \label{tab:GEOvsLEO}
    \begin{tabular}{l c c}\hline
        \textbf{Scenario} & \textbf{\acs{GEO}} & \textbf{\acs{LEO}} \\\hline
        Carrier frequency $f_0$ & \multicolumn{2}{c}{$2.2\GHz$} \\\hline
        Carrier wavelength $\lambda$ & \multicolumn{2}{c}{$13.63\cm$} \\\hline
        Orbital height $h_\mathsf{sat}$ & $35,786\km$ & $550\km$ \\\hline
        Array element spacing $\delta/\lambda$ & $4.5$ & $0.6$ \\\hline
        Number of array elements $N$ & $49$ & $625$ \\\hline
        Number of satellites $S$ & $289$ & $9$ \\\hline
        Per-satellite transmit power $P$ & $342\Watt$ & $91.5\Watt$ \\\hline
        Allocated service bandwidth $B_\mathsf{tot}$ & \multicolumn{2}{c}{$60\MHz$} \\\hline
        Number of colors $C$ & \multicolumn{2}{c}{$3$} \\\hline
        Per-color symbol rate $\Gamma_\mathsf{s}$ & \multicolumn{2}{c}{$18.18\MBaud$} \\\hline
        Minimum \acs{UT} distance $d_{\rm min}$ & $70\km$ & $40\km$ \\\hline
        \acs{UT} cardinality $K$ & $[1000, 2000]$ & $[200, 1200]$ \\\hline
        \acs{UT} minimum elev. angle $\varepsilon_{\rm min}$ & $75^\circ$ & $40^\circ$ \\\hline
    \end{tabular}
\end{table}

\subsection{Signal model and performance metrics}\label{results:signalModel}

We consider the downlink with $C$-color frequency reuse, where $K$ \acp{UT} share the same time-frequency resource within one color. Let $\mathbf{h}_k\in\mathbb{C}^{M}$ and $\widetilde{\mathbf{w}}_k\in\mathbb{C}^{M}$ denote the channel and precoding vectors for \ac{UT} $k$. For independent unit-power data symbols and treating multiuser interference as noise, the \ac{SINR} is
\begin{align}
\label{eq:sinr}
\gamma_k=
\frac{|\mathbf{h}_k^\Htran\widetilde{\mathbf{w}}_k|^2}
{\sum_{i\neq k}|\mathbf{h}_k^\Htran\widetilde{\mathbf{w}}_i|^2+1}.
\end{align}
The channel vectors are normalized by the receiver noise standard deviation and incorporate the link-budget parameters of \cite[\tablename~I]{RdG_2025}, including \ac{UT} $G/T$ and polarization, fading, and demodulation losses considered in \cite[\tablename~I]{RdG_2025}.

For each \ac{UT} deployment, we construct $\widetilde{\mathbf{H}}=[\widetilde{\mathbf{h}}_1,\ldots,\widetilde{\mathbf{h}}_K]\in\mathbb{C}^{M\times K}$ using the exact \ac{NF}, \ac{FF}, or \ac{sFF} model of \sectionname~\ref{sec:channel}. The unnormalized \ac{MF} and \ac{MMSE} precoders are $\widetilde{\mathbf{W}}^\mathsf{MF}=\widetilde{\mathbf{H}}$ and $\widetilde{\mathbf{W}}^\mathsf{MMSE} =\widetilde{\mathbf{H}}(\widetilde{\mathbf{H}}^\Htran\widetilde{\mathbf{H}}
+\Lambda \mathbf{I}_K)^{-1}$
%
%
respectively, where $\Lambda = {K\cdot C}/({S\cdot P})$ and 
$P$ is the per-satellite transmit power, accounting for payload losses and amplifier back-off. The final precoder is obtained by applying the \acl{FPAC} operator, which enforces equal per-element power budgets and the total power constraint, as detailed in \cite{RdG_2025}.

\begin{remark}[Practical relevance of precoding mismatch]
In distributed satellite systems, the low \acl{SNR} of individual radiating elements can make direct estimation of all channel coefficients impractical. Precoding therefore often relies on channel reconstruction from user-location and satellite-state information, using an assumed propagation model. The distinction between the reconstructed channel $\widetilde{\mathbf{h}}_k$ and the propagation channel $\mathbf{h}_k$ thus reflects a practical design issue. If reconstruction uses a global \ac{FF} approximation while users lie in the formation's radiative \ac{NF}, a modeling mismatch remains even with perfect geometric information. Our comparison isolates the performance penalty caused by this operationally relevant mismatch from that due to parameter-estimation errors.
\end{remark}


The spectral efficiency of \ac{UT} $k$ is $\eta_k=\Phi(\gamma_k)$, where $\Phi(\cdot)$ is the staircase \ac{SINR}-to-spectral-efficiency mapping used for \ac{3GPP} \acl{NR} link adaptation \cite{3GPP_TR-38.811}. If $\gamma_k$ falls below the lowest supported threshold, $\eta_k=0$ and the link is declared in \emph{outage}. We evaluate the \emph{average aggregate network throughput} and \emph{outage probability}:
\begin{align}
\label{eq:kpi}
\overline{T}
=C \,\Gamma_\mathsf{s} \mathbb{E}\left\{\sum\nolimits_{k=1}^{K}\eta_k\right\}
\end{align}
and
\begin{align}
P_\mathsf{out}
=\frac{1}{K} \mathbb{E}\left\{\sum\nolimits_{k=1}^{K}
\mathds{1}\{\eta_k=0\}\right\}
\end{align}
where $\Gamma_\mathsf{s}$ is the per-color symbol rate, common to all colors, and $\mathds{1}\{\cdot\}$ denotes the indicator function. Expectations are estimated by Monte Carlo simulation over random \ac{UT} deployments. Following \cite{RdG_2025}, for each spacing $\Delta$, we report the maximum average throughput over the \ac{UT} populations considered in \tablename~\ref{tab:GEOvsLEO}.

\begin{figure}[t]
\centering
\subfigure[\acs{GEO} scenario]
{\includegraphics[width=\columnwidth]{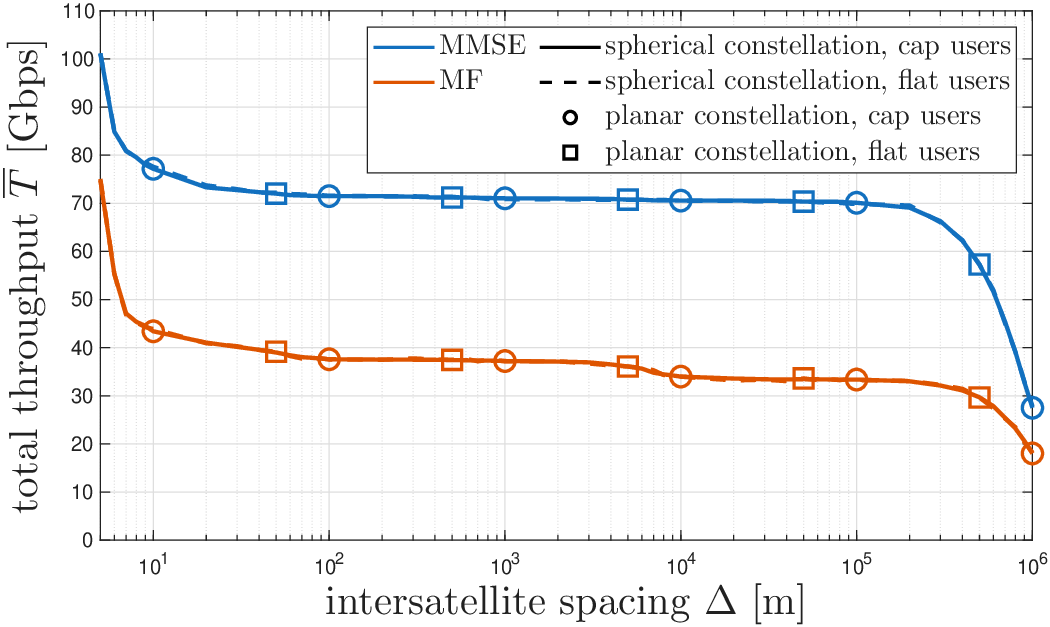}
  \label{fig:GEO_thr_vs_geometry}}

\subfigure[\acs{LEO} scenario]
{\includegraphics[width=\columnwidth]{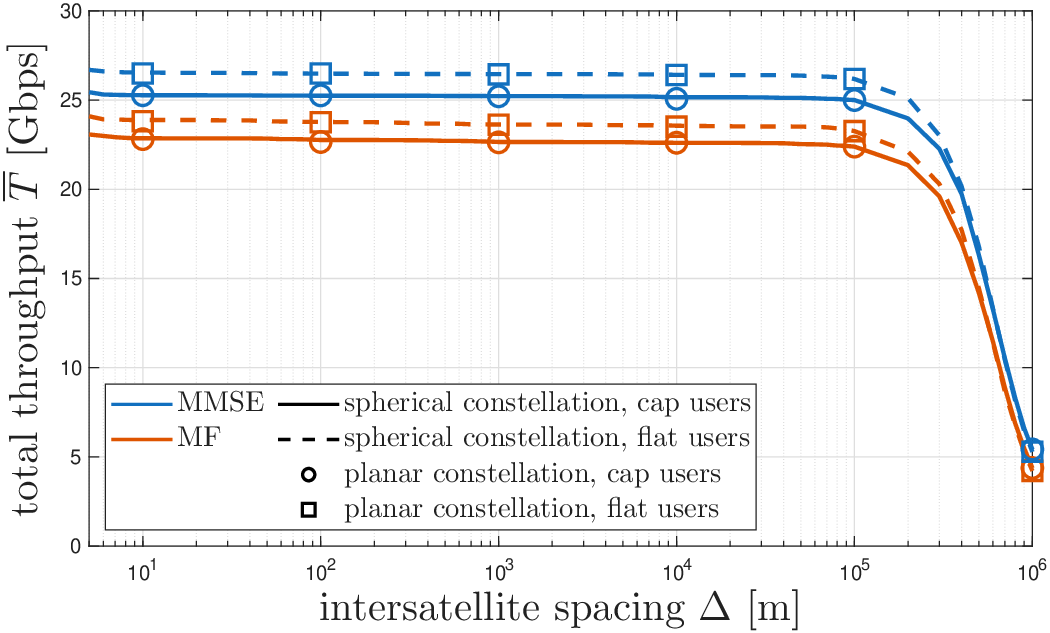}
  \label{fig:LEO_thr_vs_geometry}}
\caption{Average system throughput $\overline{T}$ as a function of $\Delta$ for different satellite and user geometries in the \subref{fig:GEO_thr_vs_geometry}~\acs{GEO} and \subref{fig:LEO_thr_vs_geometry}~\acs{LEO} scenarios.}
\label{fig:thr_vs_geometry}
\end{figure}

\subsection{Impact of Space- and User-Segment Geometry}
\label{results:geometry}

We first assess how geometric approximations affect the predicted system throughput. \figurename~\ref{fig:GEO_thr_vs_geometry} and \figurename~\ref{fig:LEO_thr_vs_geometry} compare the four combinations of conformal or planar satellite formations and spherical-cap or flat-Earth user deployments for the \ac{GEO} and \ac{LEO} scenarios, respectively. For each spacing $\Delta$, the reported throughput is maximized over the considered user populations. Channel and precoder use the same geometry and exact spherical-wave model, isolating the impact of geometric approximations. Blue and orange curves denote \ac{MMSE} and \ac{MF} precoding, respectively; the conformal formation with spherical-cap users is the reference. 

As a general observation, increasing the intersatellite spacing can reduce throughput, particularly in the \ac{GEO} scenario, consistently with \cite{RdG_2026}. Wider spacing produces a denser grating-lobe structure, which can increase multiuser interference and offset the benefit of narrower beams. Beyond this known behavior, the results of \figurename~\ref{fig:GEO_thr_vs_geometry} and \figurename~\ref{fig:LEO_thr_vs_geometry} highlights two findings:
\begin{itemize}
\item In both \ac{GEO} and \ac{LEO} scenarios, the planar \ac{FoA} approximation, widely adopted in the literature \cite{Abdelsadek_2022,Palou_2022,Wang_2025}, closely reproduces the aggregate throughput obtained with the conformal model. Thus, despite the differences in satellite positions and array orientations, formation curvature has little impact on throughput over the investigated spacings.

\item In the considered \ac{GEO} scenario, spherical-cap and flat-Earth user deployments yield nearly identical throughput. In the \ac{LEO} scenario, however, the flat-Earth approximation overestimates the maximum achievable throughput. For the same nadir-angle coverage constraint, it produces a smaller coverage area and hence a higher user density for a given population. Together with the altered satellite-to-user distances, this changes the simulated channel conditions and biases the performance prediction.
\end{itemize}

In summary, these results support the planar formation approximation in the considered scenarios, while highlighting the need to retain Earth curvature for reliable \ac{LEO} throughput predictions.

\subsection{Impact of Channel-Model Mismatch}
\label{results:precoder}

We now assess how the propagation model used to design the precoder affects throughput and outage probability. Unlike the previous comparison, all precoders are evaluated over the same exact spherical-wave channel, while their coefficients are computed using the \ac{NF}, \ac{FF}, or \ac{sFF} model. This isolates the performance loss caused by channel-model mismatch.

\begin{figure}
\centering
  \subfigure[Aggregate throughput for \acs{GEO} scenario]
    {\includegraphics[width=\columnwidth]
    {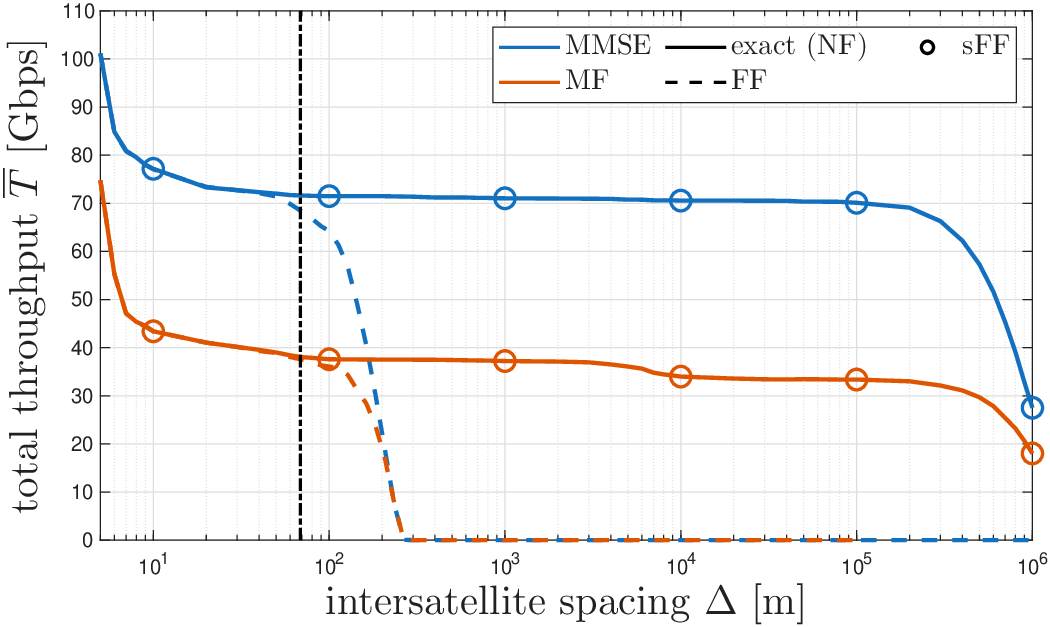}
    \label{fig:GEO_thr_vs_model}}
\\
  \subfigure[Outage probability for \acs{GEO} scenario]
    {\includegraphics[width=\columnwidth]
    {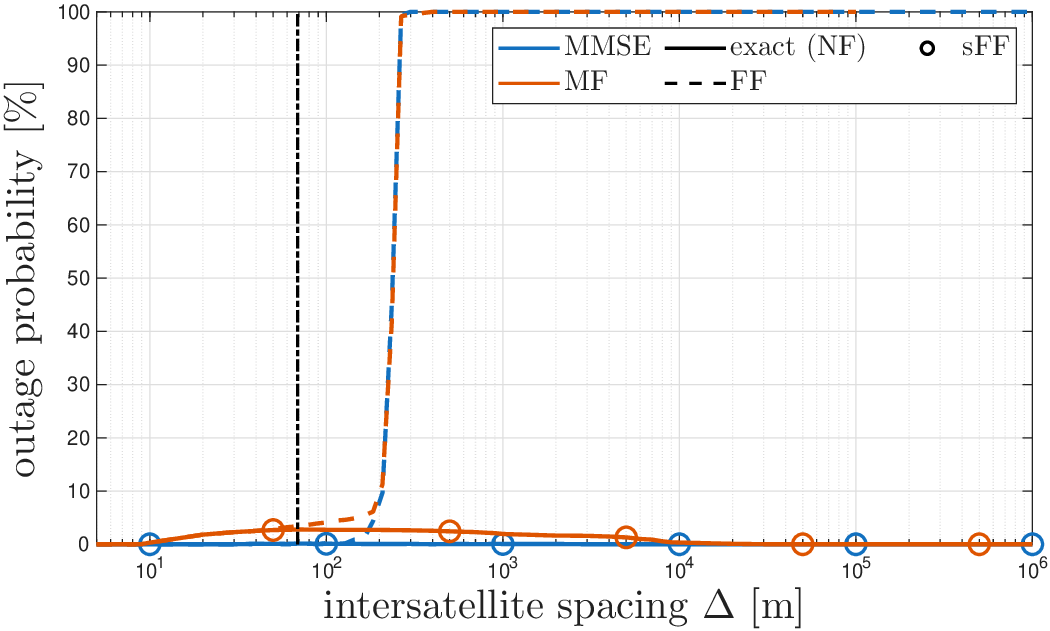}
    \label{fig:GEO_out_vs_model}}
%
\caption{Comparison of \subref{fig:GEO_thr_vs_model}~aggregate throughput and \subref{fig:GEO_out_vs_model}~outage probability as functions of $\Delta$ for different models used to compute the precoding coefficients, for both \acs{MMSE} (blue) and \acs{MF} (orange) strategies (\acs{GEO} scenario) ($\Delta < \Delta_\mathsf{F}$ corresponds to the \acs{FF} regime).}
\label{fig:GEO_vs_model}
\end{figure}

\begin{figure}
\centering
  \subfigure[Aggregate throughput for \acs{LEO} scenario]
    {\includegraphics[width=\columnwidth]
    {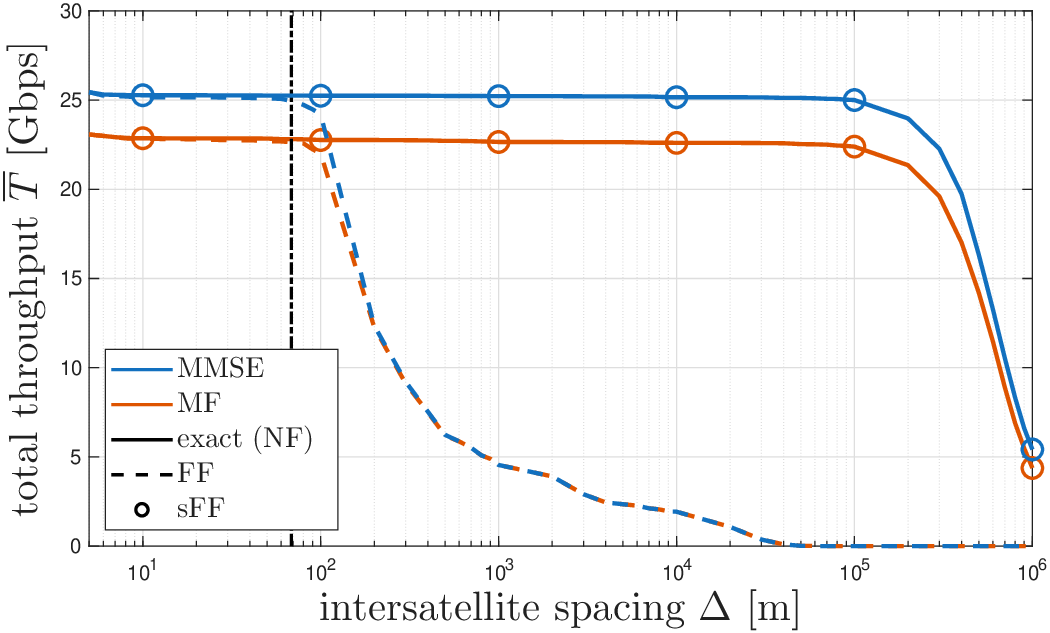}
    \label{fig:LEO_thr_vs_model}}
\\
  \subfigure[Outage probability for \acs{LEO} scenario]
    {\includegraphics[width=\columnwidth]
    {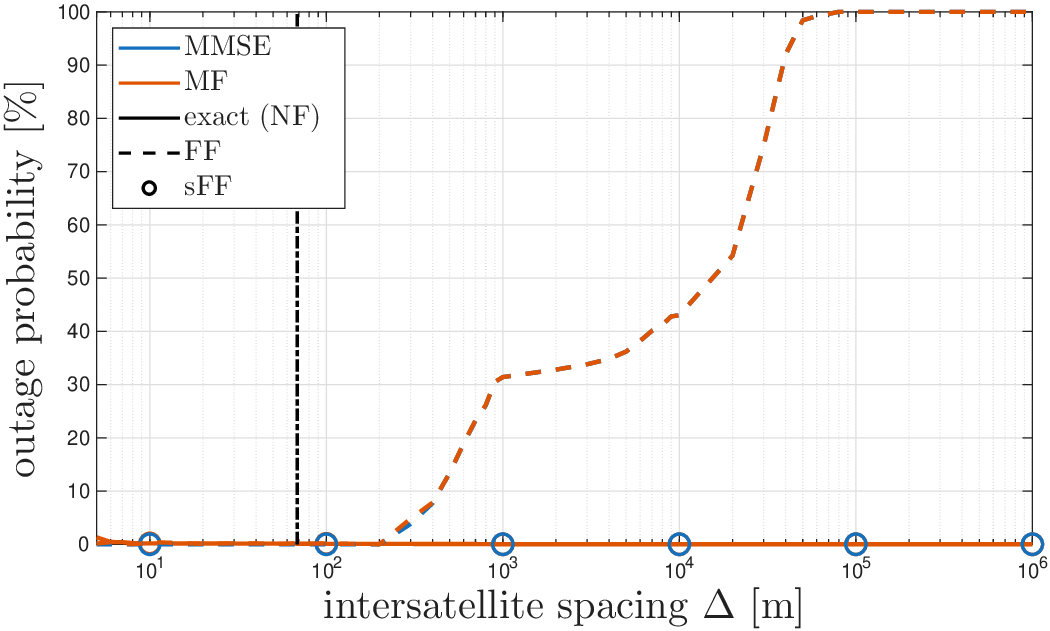}
    \label{fig:LEO_out_vs_model}}
%
\caption{Comparison of \subref{fig:LEO_thr_vs_model}~average system throughput and \subref{fig:LEO_out_vs_model}~outage probability as functions of $\Delta$ for different models used to compute the precoding coefficients, for both \ac{MMSE} (blue) and \acs{MF} (orange) strategies (\acs{LEO} scenario) ($\Delta < \Delta_\mathsf{F}$ corresponds to the \acs{FF} regime).}
\label{fig:LEO_vs_model}
\end{figure}

\figuresname~\ref{fig:GEO_vs_model} and~\ref{fig:LEO_vs_model} report the results for the \ac{GEO} and \ac{LEO} scenarios, respectively. Solid lines, dashed lines, and circular markers denote \ac{NF}-, \ac{FF}-, and \ac{sFF}-based precoding; blue and orange indicate \ac{MMSE} and \ac{MF}, respectively. Vertical dash-dotted lines mark the spacing $\Delta_\mathsf{F}$ indicating the nadir \ac{NF}-\ac{FF} boundary, which is equal to $68.8\meter$ and $67.4\meter$ for the \ac{GEO} and \ac{LEO} scenarios, respectively.

The \ac{NF}- and \ac{sFF}-based approaches closely agree in both aggregate throughput (\figuresname~\ref{fig:GEO_thr_vs_model} and \ref{fig:LEO_thr_vs_model}) and outage probability (\figuresname~\ref{fig:GEO_out_vs_model} and \ref{fig:LEO_out_vs_model}) for both scenarios, consistently with the beam-pattern analysis in \sectionname~\ref{results:beamPattern}.\footnote{For the constellation of \acp{FoA} discussed in \sectionname~\ref{spaceSegment:architecture}, this agreement requires each locally approximated unit to operate in its own \ac{FF}. Satellite-level stacking can therefore remain accurate even when cluster-level stacking fails.}

With \ac{FF}-based precoding, performance remains similar to the other approaches when $\Delta\ll\Delta_\mathsf{F}$, but degrades substantially when $\Delta\gg\Delta_\mathsf{F}$. The interpretation of these curves is important: the propagation channel is generated using the exact \ac{NF} model in every case, while only the \emph{precoder} uses the \ac{FF} approximation. The low throughput and high outage of the dashed curves therefore represent the simulated performance under mismatched precoding, rather than merely an inaccurate performance prediction. At sufficiently large spacings, the loss of coherent combining gain drives the aggregate throughput to zero and the outage probability to $100\%$ in both scenarios.

In summary, global \ac{FF} precoding can severely degrade performance when users lie in the formation's radiative \ac{NF}. This modeling issue remains insufficiently explored in satellite \ac{D2D} communications, despite related findings in terrestrial systems (e.g., \cite{Bacci_2023wcl}).

\subsection{Model Validity and Wideband Considerations}
\label{results:limitations}

The spacing range in \figuresname~\ref{fig:GEO_vs_model} and \ref{fig:LEO_vs_model} extends beyond practical operating conditions to illustrate the behavior of the models at very large apertures. Two geometric limitations should therefore be considered. First, the coverage condition of \sectionname~\ref{model:userSegment} applies the minimum elevation $\varepsilon_{\rm min}$ to the formation as a whole, which is accurate only while the formation is compact: for the \ac{LEO} system, the elevation under which the \emph{outermost} satellite is seen by the sub-satellite \ac{UT} falls to $\varepsilon_{\rm min}=40^\circ$ at $\Delta\approxeq440\km$, and to $16.1^\circ$ at $\Delta=1{,}000\km$, so that beyond roughly $440\km$ the individual satellites no longer meet the constraint that defines the service. Second, the \ac{FF} model of \sectionname~\ref{channel:ff} rests on $\max_s\theta_s\ll1$: for the \ac{GEO} system that ratio remains below $0.1$ only up to $\Delta\approxeq370\km$, beyond which the approximation $\boldsymbol{\nu}_s\approxeq\hat{\mathbf{z}}$ degrades and contributes to the \ac{FF} loss alongside the wavefront error. Both limits lie orders of magnitude above the Fraunhofer boundary $\Delta_\mathsf{F}$, and therefore do not affect the conclusions drawn around it; they do mean that the last decade of each curve is to be read as asymptotic rather than as an operating point.

Finally, the analysis is performed at the carrier frequency $f_0$ and is therefore unaffected by aperture-induced frequency selectivity; the reported \acp{KPI}, however, assume coherent beamforming across the whole signal band. This becomes restrictive as $\Delta$ grows: the differential propagation delay reaches $3.7\mus$ for the \ac{GEO} formation at $\Delta=\Delta_\mathsf{F}$, and $6.7\mus$ for the \ac{LEO} one at $\Delta=\Delta_2=1\km$, comparable with the \ac{NR} normal cyclic prefix. Two mitigations are available:
\begin{enumerate*}[label=(\emph{\roman*})]
\item With \acl{OFDM}, beamforming coefficients can be computed \emph{per subcarrier}, or per subcarrier group, to mitigate beam squint, while the cyclic prefix accommodates the residual differential delay. This requires frequency-dependent weights and increases on-board processing \cite{SatNex_YB};
\item When the residual delay exceeds the cyclic prefix, additional delay compensation is required; true-time-delay beamforming, as analyzed in \cite{Bacci_2024}, offers one solution at increased payload complexity.
\end{enumerate*}

\section{Conclusions}
\label{sec:conclusion}

This paper assessed how geometric and propagation approximations affect multi-antenna precoding in \ac{FoA}-based direct-to-device satellite systems. We combined spherical models of the space and user segments with exact spherical-wave propagation, and evaluated geometric simplifications and channel-model mismatch through throughput and outage analyses. For the investigated scenarios, planar and conformal formations yield nearly identical throughput, whereas the flat-Earth user model overestimates \ac{LEO} performance. These results support simplifying the formation geometry while retaining Earth curvature where it materially affects user coverage and deployment. 

The dominant modeling issue is wavefront curvature across the distributed aperture. The Fraunhofer distance depends on the formation's extent, determined by satellite number, arrangement, and spacing, as well as the carrier wavelength. For the considered S-band systems, nadir users enter the radiative near field at spacings of approximately $70\meter$ for both the geostationary formation and the low-Earth-orbit system. Although the large propagation distances of satellite links may suggest that a single plane-wave approximation suffices across the entire formation, using it to design the precoder can severely degrade performance, leading to full outage in the investigated configurations. Conversely, approximating the wavefront only within each satellite array closely matches the exact spherical-wave benchmark, provided that satellite-center-to-user distances are retained exactly and users remain in each array's far field. In hierarchical formations, this condition must hold for each unit to which the local approximation is applied.



These findings have broader implications for satellite communication tasks that rely on parametric channel models. Neglecting relevant geometric or wavefront curvature introduces model mismatch even when the underlying parameters are known exactly. The resulting errors can compromise channel reconstruction and subsequent signal processing, as demonstrated here by the performance loss under mismatched precoding. Similar modeling considerations apply beyond communications, including satellite-based localization.

\appendix

\subsection{Satellite placement for the conformal configuration}\label{app:conformal_satellite}

To illustrate the criterion sketched in \sectionname~\ref{model:spaceSegment} to compute $\mathbf{r}_s$, we first associate with each satellite an auxiliary point $\tilde{\mathbf{r}}_s=\left[\tilde{x}_s=i_s R\varphi, \tilde{y}_s=j_s R\varphi, \tilde{z}_s=0\right]^\Ttran$ on the plane tangent to the sphere at the \ac{FoA} center $\mathbf{O}$, where the definitions \eqref{eq:phi}, \eqref{eq:i}, and \eqref{eq:j} are used. Hence, the auxiliary tangent-plane grid has horizontal and vertical spacing $R\varphi$. Note that, by construction, $\|\tilde{\mathbf{r}}_s\|=R\varphi\sqrt{i_s^2+j_s^2}=\ell_s$, where $\ell_s$ is defined in \eqref{eq:ell_s}. 

\begin{figure}[t]
  \centering
   \includegraphics[width=\columnwidth]{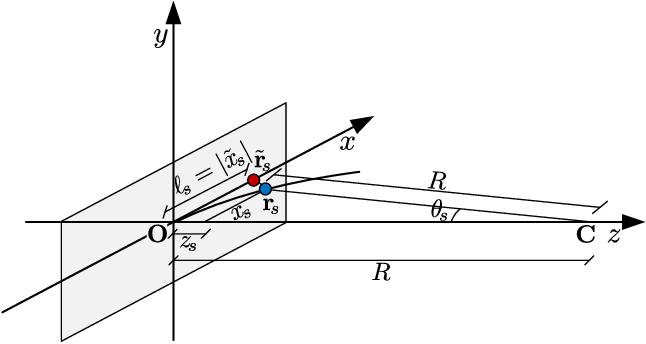}
  \caption{Mapping of the auxiliary tangent-plane point $\tilde{\mathbf{r}}_s$ onto the spherical surface for the particular case $j_s=0$, with geodesic distance $\ell_s$.}
  \label{fig:projection}
\end{figure}

To get $\mathbf{r}_s$, we then need to map $\tilde{\mathbf{r}}_s$ onto the spherical surface so as to preserve both its radial direction in the tangent plane and its distance from $\mathbf{O}$, with the latter interpreted as a geodesic distance on the sphere. To visualize the situation, \figurename~\ref{fig:projection} depicts a simplified case with $\tilde{\mathbf{r}}_s=\left[\tilde{x}_s, \tilde{y}_s=0, \tilde{z}_s=0\right]^\Ttran$ (indicated by the red dot, corresponding to a satellite belonging to the central row of the \ac{FoA}, with $j_s=0$). The geodesic distance with respect to $\mathbf{O}$ is equal to the arc $\ell_s=R\theta_s$, where $\theta_s$ denotes the central angle subtended by the maximum-circle arc joining $\mathbf{O}$ and $\mathbf{r}_s$. As a consequence, $\theta_s=\ell_s/R$. The coordinates $\mathbf{r}_s=\left[x_s,y_s=0,z_s\right]^\Ttran$ (represented in \figurename~\ref{fig:projection} by the blue dot, lying on the sphere surface and forming an arc with length $\ell_s$ with respect to $\mathbf{O}$) can be obtained using elementary trigonometry: $x_s=R\cdot\sin\theta_s=R\cdot\sin\left(\ell_s/R\right)$ and $z_s=R\left(1-\cos\theta_s\right)=R\left[1-\cos\left(\ell_s/R\right)\right]$, which correspond to \eqref{eq:conformal_satellite_x}-\eqref{eq:conformal_satellite_z} when $j_s=0$.

\begin{figure}[t]
  \centering
  \includegraphics[width=0.8\columnwidth]{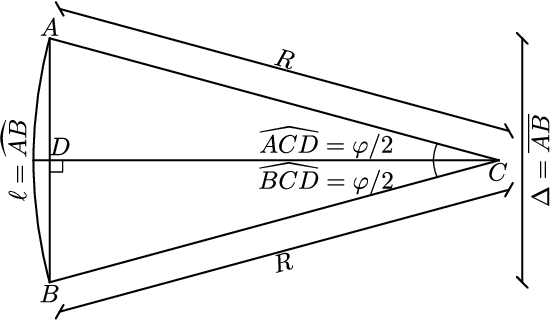}  
  \caption{Relation between the arc $\ell$ and the chord $\Delta$.}
  \label{fig:arc}
\end{figure}

To compute the angle $\varphi$ \eqref{eq:phi} used in \eqref{eq:ell_s}, we can focus on two adjacent satellites along the central row, whose coordinates are represented in \figurename~\ref{fig:arc} by points $A$ and $B$, lying on a circle with radius $R$. When the chord $\overline{AB}$ is equal to the spacing $\Delta$, the circular arc $\ell$ is 
\begin{align}\label{eq:arc}
  \ell=R\cdot\varphi>\Delta,
\end{align}
where $\varphi=\widehat{ACB}$. We can see that $\Delta/2=R\cdot\sin(\varphi/2)$, which leads to \eqref{eq:phi}.

The construction of the spherical grid, and in particular  the derivation of the locations of noncentral satellites with $\left(i_s, j_s\right)\neq(0,0)$, is based on the general principle of giving  $\tilde{\mathbf{r}}_s$ the same direction as ${\mathbf{r}}_s$ on the tangent plane. To this end, we define the unit vector $\hat{\mathbf{t}}_s=\left.\left[i_s, j_s, 0\right]^\Ttran\right/\sqrt{i_s^2+j_s^2}$, which identifies the direction of the satellite $s$ in the tangent plane with respect to $\mathbf{O}$. For the central satellite, $\rho_s=0$, and the direction is irrelevant, as the point stays at the origin. As in the example depicted in \figurename~\ref{fig:projection}, the projection of $\mathbf{r}_s$ onto the tangent plane has magnitude $R\cdot\sin\left(\ell_s/R\right)$ and direction $\hat{\mathbf{t}}_s$, whereas its height above the tangent plane is $R\cdot\left[1-\cos\left(\ell_s/R\right)\right]$. Therefore, $\mathbf{r}_s=\hat{\mathbf{t}}_s\cdot R\sin\left(\ell_s/R\right)+\hat{\mathbf{z}}\cdot R\cdot\left[1-\cos\left(\ell_s/R\right)\right]$, where $\hat{\mathbf{z}}=\left[0, 0, 1\right]^\Ttran$. Expanding the above vector expression directly yields \eqref{eq:conformal_satellite_x}-\eqref{eq:conformal_satellite_z}.

\subsection{Antenna placement for the conformal configuration}\label{app:conformal_antenna}
\begin{figure}[t]
  \centering
    \includegraphics[width=0.8\columnwidth]{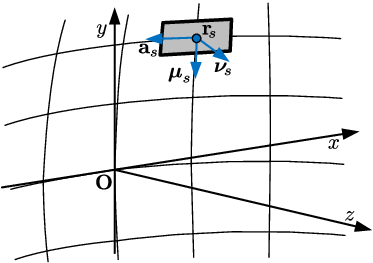}
  \caption{Local orthonormal frame at the satellite center $\mathbf{r}_s$, with the satellite array lying on the plane tangent to the spherical surface of radius $R$.}
  \label{fig:tangent_plane}
\end{figure}
The $N$ radiating elements of each satellite array centered in $\mathbf{r}_s$ are located on a plane tangent to the sphere at $\mathbf{r}_s$. Such plane has a normal vector $\boldsymbol{\nu}_s$, equal to the radial direction, as indicated in \figurename~\ref{fig:tangent_plane}. To compute it, we observe that $\boldsymbol{\nu}_s=-{\mathbf{r}_s^\prime}/{\|\mathbf{r}_s^\prime\|}$, where $\mathbf{r}_s^\prime=\mathbf{r}_s-\mathbf{C}$, represents the satellite $s$'s coordinates in the Cartesian reference system centered in the sphere center $\mathbf{C}=\left[0, 0, R\right]^\Ttran$, so that $\|\mathbf{r}_s^\prime\|=R$. The minus sign in \eqref{eq:normal_direction} selects the inward radial direction, pointing from the satellite toward the sphere center $\mathbf{C}$ (i.e., the array boresight direction, with positive $z$-component $\cos\theta_s$, as drawn in \figurename~\ref{fig:tangent_plane}). By exploiting the quantities introduced in \sectionname~\ref{model:spaceSegment}, we can get \eqref{eq:normal_direction} (note that, by construction, $\|\boldsymbol{\nu}_s\|=1$).

The tangent-plane constraint uniquely determines the array normal $\boldsymbol{\nu}_s$, but leaves one rotational degree of freedom about this normal. We remove this ambiguity by orienting the local array axes along the azimuthal and meridional directions, denoted by $\mathbf{a}_s$ and $\boldsymbol{\mu}_s$, respectively.

To define a local orthonormal basis on the tangent plane, we first introduce the azimuthal direction $\mathbf{a}_s$, indicated in \figurename~\ref{fig:tangent_plane} and computed as in \eqref{eq:azimuthal_direction}. Its orthogonality with respect to $\boldsymbol{\nu}_s$ is guaranteed by $\boldsymbol{\nu}_s^\Ttran\mathbf{a}_s=0$, while $\|\mathbf{a}_s\|=1$ follows from \eqref{eq:hat}.

To determine the desired tangent plane, we need to identify two independent vectors lying inside it and orthogonal to $\boldsymbol{\nu}_s$. One such vector is given by the azimuthal direction $\mathbf{a}_s$, indicated in \figurename~\ref{fig:tangent_plane} and computed as \eqref{eq:azimuthal_direction}, whose orthogonality with respect to $\boldsymbol{\nu}_s$ is guaranteed by the dot product $\boldsymbol{\nu}_s^\Ttran \cdot \mathbf{a}_s=0$. 

The additional vector needed to identify the tangent plane can be derived by completing an orthonormal basis, and thus obtaining the meridional vector $\boldsymbol{\mu}_s$ (\figurename~\ref{fig:tangent_plane}) using the cross product $\boldsymbol{\mu}_s=\boldsymbol{\nu}_s \times \mathbf{a}_s$, which yields \eqref{eq:meridional_direction}. 

To place the original array template, whose coordinates are $\mathbf{r}_n=\left[i_n\delta, j_n\delta, 0\right]^\Ttran$, where $i_n$ and $j_n$ are defined as in \eqref{eq:i_n}, on the tangent plane centered around the satellite coordinates $\mathbf{r}_s=\left[x_s, y_s, z_s\right]^\Ttran$, computed using \eqref{eq:conformal_satellite_x}-\eqref{eq:conformal_satellite_z}, we need to consider the vectors $\mathbf{a}_s$, $\boldsymbol{\mu}_s$, and $\boldsymbol{\nu}_s$ as the local $x$, $y$, and $z$ axes, respectively. In other words, the position $\mathbf{u}_{s,n}$ of the $n$th antenna in the $s$th satellite array becomes \eqref{eq:conformal_antenna}. Finally, note that, for the central satellite, the azimuthal direction is undefined. Since the array normal is $\boldsymbol{\nu}_s=\left[0,0,1\right]^\Ttran$, we conventionally set $\mathbf{a}_s=\left[1,0,0\right]^\Ttran$ and $\boldsymbol{\mu}_s=\left[0,1,0\right]^\Ttran$, yielding $\mathbf{P}_s=\mathbf{I}_3$.

\subsection{Fraunhofer distance for planar and conformal satellite configurations}\label{app:fraunhofer}

Let us derive the Fraunhofer distance for an observer located at $\mathbf{m}=[0,0,z]^\Ttran$ along the nadir direction, with respect to an \ac{FoA} using the conformal placement detailed in \sectionname~\ref{model:spaceSegment}. The distance of $\mathbf{m}$ from the radiating element $n$ of satellite $s$ is equal to 
\begin{align}\label{eq:distance_conf}
  d_{s,n}(z)=\|\mathbf{m}-\mathbf{u}_{s,n}\|=\sqrt{x_{s,n}^2+y_{s,n}^2+(z-z_{s,n})^2}.
\end{align}
The \ac{FF} (plane-wave) approximation replaces such exact distance with its projection along the observation direction $\hat{\mathbf{z}}=[0,0,1]^\Ttran$:
\begin{align}\label{eq:ff_conf}
  d_{s,n}^\mathsf{FF}(z)=z-\hat{\mathbf{z}}\cdot\mathbf{u}_{s,n}=z-z_{s,n}>0
\end{align}
where $-z_{s,n}$ accounts for the first-order path-length correction introduced by the element $n$ of the satellite $s$ along the observation direction, and where the inequality follows from the customary assumption $z>\max_{s,n}z_{s,n}$. This means that this introduces a phase error equal to
\begin{align}\label{eq:phase_error_conf}
  \Delta\phi_{s,n}(z)=\kappa\left[d_{s,n}(z)-d_{s,n}^\mathsf{FF}(z)\right]= \kappa\cdot \epsilon_{s,n}(z),
\end{align}
where $\kappa=2\pi/\lambda$ is the wave number, and where
\begin{align}\label{eq:path_error_conf}
  \epsilon_{s,n}(z)&\triangleq d_{s,n}(z)-d_{s,n}^\mathsf{FF}(z)\nonumber\\
  &=
  \sqrt{x_{s,n}^2+y_{s,n}^2+\left(z-z_{s,n}\right)^2}-(z-z_{s,n})
\end{align}
is the path-length error. We  define the Fraunhofer distance $z_\mathsf{F}$ for the conformal configuration as the minimum distance such that
\begin{align}\label{eq:fraunhofer_conf}
  \max_{s,n} \epsilon_{s,n}\left(z_\mathsf{F}\right) = \frac{\pi}{8\kappa}=\frac{\lambda}{16},
\end{align}
corresponding to a maximum phase error $\pi/8$. By letting $\epsilon_{s,n}(z)=\lambda/16$, we get
\begin{align}\label{eq:path_error_conf_2}
  z=z_{s,n}+\frac{8}{\lambda}\left(x_{s,n}^2+y_{s,n}^2\right)-\frac{\lambda}{32}.
\end{align}
As $\epsilon_{s,n}(z)$ is monotonically nonincreasing with $z$, using \eqref{eq:path_error_conf_2} in \eqref{eq:fraunhofer_conf} we obtain
\begin{align}\label{eq:fraunhofer_conf_2}
  z^\star_\mathsf{F}=\max_{s,n}\left\{z_{s,n}+\frac{8}{\lambda}\left(x_{s,n}^2+y_{s,n}^2\right)\right\}-\frac{\lambda}{32}.
\end{align}
in which the superscript $^\star$ is used to emphasize that this is the numerically exact solution.

We derive now simple, closed-form approximations of such expression. First, let us assume that the individual satellite-array size is negligible with respect to the overall \ac{FoA} extent, i.e., $L \ll (\sqrt{S}-1)\Delta$. Under this assumption, the displacement of each radiating element with respect to its satellite center provides a negligible contribution to the overall aperture extent and, consequently, to the maximum path-length error. We can therefore approximate $\mathbf{u}_{s,n}\approxeq\mathbf{r}_s$, meaning that
\begin{align}\label{eq:distance_conf_approx}
  d_{s,n}(z)&\approxeq\|\mathbf{m}-\mathbf{r}_s\|=\sqrt{x_s^2+y_s^2+\left(z-z_s\right)^2}\\
  d_{s,n}^\mathsf{FF}(z)&\approxeq z-\hat{\mathbf{z}}\cdot\mathbf{r}_{s}=z-z_s\\
  \epsilon_{s,n}(z)&\approxeq \sqrt{x_s^2+y_s^2+\left(z-z_s\right)^2}-(z-z_s).
\end{align}
Then, let us consider that $z_\mathsf{F}$ is much larger than the formation size, in particular $z_\mathsf{F}\gg\|\mathbf{r}_s\|$ for any $s$. Under this hypothesis, both $z_\mathsf{F}\gg\sqrt{x_s^2+y_s^2}$ and $z_\mathsf{F}\gg z_s$ hold, so that

\begin{align}\label{eq:path_error_conf_3}
  \epsilon_{s,n}\left(z_\mathsf{F}\right)\approxeq (z_\mathsf{F}-z_s)\left(\sqrt{1+\frac{x_s^2+y_s^2}{(z_\mathsf{F}-z_s)^2}}-1\right)\approxeq
  \frac{x_s^2+y_s^2}{2z_\mathsf{F}},
\end{align}
where the last step uses $\sqrt{1+a}\approxeq1+a/2$ for $a\ll1$. Using \eqref{eq:path_error_conf_3} in \eqref{eq:fraunhofer_conf}, we get
\begin{align}\label{eq:fraunhofer_conf_3}
  z_\mathsf{F}\approxeq \frac{8}{\lambda} \max_{s} \left(x_{s}^2+y_{s}^2\right).
\end{align}
Relation \eqref{eq:fraunhofer_conf_3} holds for any formation, since the maximization runs over the occupied grid nodes only; its closed form, however, depends on \emph{which} nodes are occupied (\sectionname~\ref{spaceSegment:conformal}). For a \emph{fully populated} $\sqrt{S}\times\sqrt{S}$ grid, and for the spherical caps of practical interest, the maximum transverse distance is attained at the four corner satellites, i.e., when $i_s=j_s=\pm(\sqrt{S}-1)/2$. Using \eqref{eq:conformal_satellite_x}-\eqref{eq:conformal_satellite_z}, the relation \eqref{eq:fraunhofer_conf_3} becomes \eqref{eq:fraunhofer_conf_main}. 
%
\begin{figure}[t]
  \centering
    \includegraphics[width=\columnwidth]{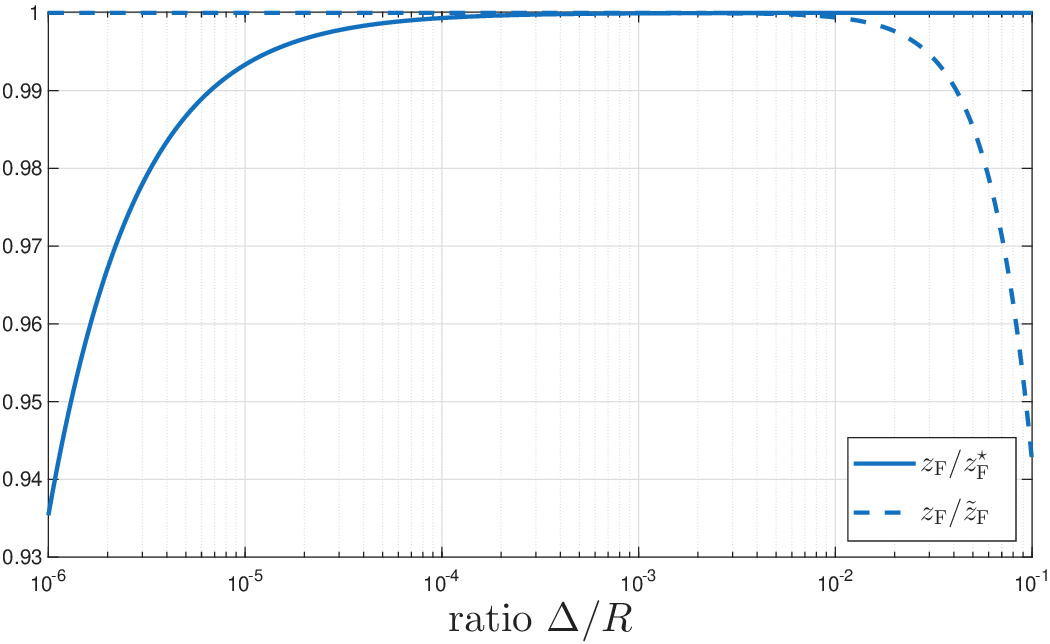}
  \caption{Ratio of the exact vs. approximate Fraunhofer distance as a function of the ratio $\Delta/R$ for a \acs{LEO} constellation ($h_\mathsf{sat}=550\km$) of $S=49$ satellites, with $N=625$ elements spaced by $\delta=0.6\lambda=8.18\cm$.}
  \label{fig:fraunhofer_conf_vs_planar}
\end{figure}

To compare the relations derived above, let us use some numerical examples, considering the configuration introduced in \sectionname~\ref{model:spaceSegment} to plot \figurename~\ref{fig:model_conf_vs_planar}. In particular, the solid and the dashed lines in \figurename~\ref{fig:fraunhofer_conf_vs_planar} report the ratios $z_\mathsf{F}/z^\star_\mathsf{F}$ and $z_\mathsf{F}/\tilde{z}_\mathsf{F}$, respectively, where $z_\mathsf{F}$, $\tilde{z}_\mathsf{F}$, and $z^\star_\mathsf{F}$ are computed using \eqref{eq:fraunhofer_conf_main}, \eqref{eq:fraunhofer_plan_main}, and \eqref{eq:fraunhofer_conf_2}, respectively, as functions of the ratio $\Delta/R$, where $R=R_\mathsf{E}+h$, with $R_\mathsf{E}=6,371\km$ and $h_\mathsf{sat}=550\km$, considering a constellation with $S=49$ satellites, each hosting an \ac{UPA} with $N=625$ elements, with interelement spacing $\delta=0.6\lambda$, with $\lambda\approxeq13.6\cm$ (obtained for a carrier frequency $f_0=2.2\GHz$, and hence $L=(\sqrt{N}-1)\delta\approxeq 1.96\meter$). 

As expected, when $\Delta$ is very small (in the minimum value considered in this numerical example, $\Delta=10^{-6}R\approxeq6.92\meter\approxeq3.5L$), it becomes comparable with the satellite size $L$, so that the approximation $L \ll (\sqrt{S}-1)\Delta$ is no longer valid and the expression for $\tilde{z}_\mathsf{F}$ is inaccurate.  However, as observed in \sectionname~\ref{channel:fraunhofer}, in this case the Fraunhofer distance is much smaller than $h_\mathsf{sat}$, and all \acp{UT} are located in the \ac{FF} region, which means that this approximation inaccuracy does not impact the analysis conducted in \sectionname~\ref{sec:results}. As soon as $\Delta$ increases (e.g., when $\Delta=10^{-4}R\approxeq692.1\meter$), all values become practically indistinguishable. When $\Delta$ becomes extremely large (e.g., when $\Delta=10^{-1}R\approxeq692.1\km$), the spherical-based distances (both the approximated and the exact ones, as confirmed by the solid line) diverge from the planar-based one (dashed line). The physical interpretation is that the spherical curvature pushes outer satellites closer to the nadir observer (by the sagitta $z_{s}$), thus altering the quadratic error relative to the planar case. However, bending the system reduces the overall aperture seen by a remote observer with respect to the planar case, thus reducing the Fraunhofer distance (in other words, the effect of reducing the seen aperture is dominating over putting the satellites closer to the observer). Nevertheless, it is worth noting that such values of $\Delta$ represent extreme cases which are not relevant in practical systems, as outlined in \sectionname~\ref{channel:fraunhofer}. As a conclusion, in all cases of practical interest, we can reasonably use the approximation \eqref{eq:fraunhofer_plan_main}. Similarly to \figurename~\ref{fig:model_conf_vs_planar}, the conclusions drawn for the \ac{LEO} case apply to the \ac{GEO} scenario as well.


\bibliographystyle{IEEEtran}
\bibliography{IEEEabrv, refs, refs_v210_extra}


\end{document}